\documentclass[preprint,12pt]{elsarticle}

\usepackage[outdir=./plots/epstopdf/]{epstopdf}
\usepackage{bm}
\usepackage{amsmath}
\usepackage{graphicx}
\usepackage{subcaption}
\usepackage{multirow} 
\usepackage{makecell} 
\usepackage{pdflscape}
\usepackage{tabularx} 
\usepackage{overpic}
\usepackage{placeins}
\usepackage{caption}
\usepackage[ruled,vlined]{algorithm2e}
\usepackage[justification=centering]{caption}
\usepackage{algpseudocode}
\usepackage{xcolor}

\usepackage{amssymb}
\usepackage{amsthm}

\usepackage{lineno}

\DeclareMathOperator*{\argmin}{arg\,min}
\DeclareMathOperator*{\argmax}{arg\,max}

\biboptions{sort&compress}

\journal{To be determined}

\begin{document}

\begin{frontmatter}



\title{Bayesian improved cross entropy method with categorical mixture models}

\author{Jianpeng Chan}
\author{Iason Papaioannou}
\author{Daniel Straub}
\address{Engineering Risk Analysis Group, Technische Universit{\"a}t M{\"u}nchen, Arcisstr. 21, 80290 M{\"u}nchen, Germany}

\begin{abstract}
We employ the Bayesian improved cross entropy (BiCE) method for rare event estimation in static networks and choose the categorical mixture as the parametric family to capture the dependence among network components. 
At each iteration of the BiCE method, the mixture parameters are updated through the weighted maximum a posteriori (MAP) estimate, which mitigates the overfitting issue of the standard improved cross entropy (iCE) method through a novel balanced prior, and we propose a generalized version of the expectation-maximization (EM) algorithm to approximate this weighted MAP estimate. 
The resulting importance sampling distribution is proved to be unbiased. 
For choosing a proper number of components $K$ in the mixture, we compute the Bayesian information criterion (BIC) of each candidate $K$ as a by-product of the generalized EM algorithm. 
The performance of the proposed method is investigated through a simple illustration, a benchmark study, and a practical application. 
In all these numerical examples, the BiCE method results in an efficient and accurate estimator that significantly outperforms the standard iCE method and the BiCE method with the independent categorical distribution. 
\end{abstract}

\begin{keyword}
network reliability assessment, Bayesian cross entropy method, categorical mixtures, Bayesian information criterion 
\end{keyword}

\end{frontmatter}

\graphicspath{{figures/}{tikz/}{plots/}{epstopdf/}}


\section{Introduction}
\label{sec: Intro}
In February 2021, three heavy winter storms swept over Texas and triggered one of the worst energy network failures in Texas state history, which soon led to a severe power, food, and water shortage. 
A conservative estimate of the property damage is over 195 billion US dollars and more than 246 (estimated) people died during this event. 
These devastating consequences highlight the need for understanding and managing the reliability of infrastructure networks. 
This requires an effective means for quantifying the probability of survival or, conversely, the probability of failure of network systems.

In this context, the network is often simplified as a graph, whose edges or/and nodes are subjected to random failure. 
The network's performance is therefore a random variable and the probability that the network cannot deliver a certain level of performance is referred to as the failure probability $p_f$. 
Mathematically, $p_f$ is defined through a performance function, $g(\cdot)$, which gives the safety margin of the network performance, and through a probabilistic input, $p_{\bm{X}}(\cdot)$, that quantifies the uncertainty of the system state $\bm{X} \triangleq [X_1,...X_d,...,X_D]^T$. $X_d$ represents the state of the $d$-th component of the network, either edge or node, and $D$ is the total number of components. 
In particular, $p_f$ reads
\begin{equation}
\label{Eq: failure probability}
p_f \triangleq \Pr \{g(\bm{X}) \leqslant 0\} = \sum_{\bm{x} \in \Omega_{\bm{X}}} \mathbb{I}\{g(\bm{X}) \leqslant 0 \} p_{\bm{X}}(\bm{x}),
\end{equation}
where $\Omega_{\bm{X}}$ is the sample space of $\bm{X}$, and $\mathbb{I} \{ \cdot \}$ represents the indicator function. 
Note that $\bm{X}$ is often discrete in the context of network reliability assessment. 
Hence, in Eq.~(\ref{Eq: failure probability}) the failure probability $p_f$ is written as a summation of the input distribution $p_{\bm{X}}(\cdot)$ over the failure domain $F \triangleq \{\bm{x} \in \Omega_{\bm{X}}: g(\bm{x}) \leqslant 0 \}$.
 
The static(or time-independent) performance of networks can often be measured by either connectivity or 'flow' \cite{Zwirglmaier&others2023}. 
For computer and communication networks, the connection among different parts of the network is of major concern, resulting in three different types of connectivity-based problems, namely the two terminals, $K$ terminals, and all terminals connectivity problems \cite{Ball&others1995}, while for road networks and food supply chains, one is primarily interested in the 'flow' that a network can deliver, e.g., the maximum flow that can be transported from A to B. 
These flow-based problems involve multi-state (even continuous) components or/and network performance and can often be regarded as an extension of the connectivity-based problems \cite{Lisnianski&Levitin2003}.
 
In Table~\ref{Table: comparison of methods for connectivity-based problems}, we summarize three state-of-art methods for solving connectivity/flow-based problems, where CB is short for the counting-based method \cite{Duenas&others2017, Paredes&others2019}, SSD is for the state-space decomposition \cite{Doulliez&Jamoulle1972, Alexopoulos1997, Li&He2002, Lim&Song2012, Paredes&others2018}, and CP is for creation process embedded methods \cite{Elperin&others1991, Hui&others2005, Murray&others2013, Vaisman&Kroese2016, Botev&others2013, Botev&others2018, Cancela&others2019, Cancela&others2022}. 
Other widely used methods include, sum of disjoint products \cite{Ball1988}, binary decision diagram \cite{Hardy&others2007}, domination theory \cite{Agrawal&Barlow1984}, and various minimal-cutsets/pathsets-based methods, e.g., \cite{Provan&Ball1984, Lin&others1995, Cancela&Khadiri1995, Cancela&others2012, Zuo&others2007}.

\begin{table}[htbp]
    \centering
    \footnotesize
    \caption{Comparison of different methods for connectivity-based problems}
    \label{Table: comparison of methods for connectivity-based problems}
    \begin{tabular}{llll}
	\hline
    & CB & SSD & CP\\  
    \hline       
    introduction & [1,2] & [3-7] & [8-13]  \\ 
    not suitable for & small comp. failure prob. & large scale network & costly $g(\cdot)$\\ 
    multi-state extension & unknown & possible & possible\\
    coherent system & needed & needed & needed\\
    error estimate & user-specific & reliability bound & relative error \\      
    \hline  
    \end{tabular}
\end{table}
For power grids and water supply systems, the 'flow' is often driven by the physical law (e.g. Kirchhoff's law for power flow) and operation strategies, and the network is not necessarily coherent. 
Hence, approaches built on the coherency assumption are not directly applicable. 
A set of methods have been proposed to solve such problems, among which sampling-based methods feature prominently. 
These include crude Monte Carlo simulation (MCS) \cite{Fishman1986a, Zio2013}, subset simulation \cite{Zio&Pedroni2008, Zuev&others2015, Jensen&Jerez2018, Chan&others2022a, Zwirglmaier&others2023}, adaptive importance sampling (IS) \cite{Kaynar&Ridder2010, Kurtz&Song2013, Chan&others2022a, Chan&others2023}, and active learning methods \cite{Cadini&others2017, Dehghani&others2021}. 
We mainly focus on the static rare event estimation for network performance in this paper, and therefore, methods for time-dependent network reliability estimation such as the probability density evolution method (PDEM) \cite{Li2020} and modern stochastic process methods \cite{Lisnianski&Levitin2003} are not included here.
 
Recently, the authors employed the improved cross entropy method (iCE) for solving network reliability problems and introduced a Bayesian approach to circumvent the overfitting issue of the standard iCE. 
The proposed method is termed Bayesian iCE (BiCE) \cite{Chan&others2023}.
Therein, the parametric model for approximating the optimal IS distribution is an independent categorical distribution and hence does not account for the dependence among components in the optimal IS distribution.
This motivates the idea of employing a more flexible categorical mixture as the parametric model within the BiCE method. 
This parametric model can be updated at each iteration of the BiCE method by the generalized EM algorithm, which is introduced in this paper to approximate the maximum a posteriori (MAP) estimate of the mixture parameters given weighed samples. 
Note that the EM algorithm for estimating the MAP of a mixture model is well known \cite{Murphy2012}; herein we develop a modified version that accounts for the sample weights. 
The major contribution of this paper is to combine this generalized EM algorithm with the BiCE method for handling a more flexible mixture parametric family. 
We find that the proposed method, termed BiCE-CM, clearly outperforms the BiCE method with a single independent categorical distribution and provides better results than the standard iCE method. 
The key ingredient of the proposed method is a balanced Dirichlet prior that does not dominate but can still correct the potentially overfitted weighted MLE in the iCE. 
A number of components $K$ in the categorical mixture is chosen adaptively through the Bayesian information criterion (BIC). 

The paper is organized as follows: 
In Sec.~2, we summarize the basic ideas of iCE, followed by a brief introduction of the categorical mixture model and its approximated inference techniques in Sec.~3. 
The BiCE method with a categorical mixture parametric family (BiCE-CM) is introduced in Sec.~4. 
The efficiency and accuracy of the proposed method are demonstrated by a set of numerical examples in Sec.~5. 
\section{Cross-entropy-based importance sampling}
\label{sec: CE based IS}
In this section, we give a brief introduction to CE-based IS \cite{DeBoer&others2005}. 
The basic idea is to choose the IS distribution from a predefined parametric family $h(\cdot; \bm{v})$ that best resembles the optimal IS distribution 
\begin{equation}
\label{Eq: optimal IS distribution}
p^*_{\bm{X}}(\bm{x}) = \frac{p_{\bm{X}}(\bm{x})\mathbb{I}\{g(\bm{x})\leq0 \}}{p_f} = p_{\bm{X}}(\bm{x}|F).
\end{equation} 
The similarity between $p_{\bm{X}}^*(\cdot)$ and $h(\cdot; \bm{v})$ is measured by the Kullback–Leibler (KL) divergence that is defined as follows:
\begin{align}
D( p_{\bm{X}}^*(\cdot), h(\cdot; \bm{v}) )&= \mathbb{E}_{p_{\bm{X}}^*} \left[ \ln \left( \frac{p_{\bm{X}}^*(\bm{X})}{ h(\bm{X}; \bm{v}) } \right) \right] \notag \\
\label{Eq: KL}
&=\mathbb{E}_{p_{\bm{X}}^*}[\ln(p_{\bm{X}}^*(\bm{X}))]-\mathbb{E}_{p_{\bm{X}}^*}[\ln(h(\bm{X}; \bm{v}))].
\end{align} 
In other words, the CE method determines the optimal parameter vector $\bm{v}^*$ in $h(\cdot;\bm{v})$ through minimizing the KL divergence in Eq.~(\ref{Eq: KL}), i.e., through solving 
\begin{align}
\bm{v}^* 
&= \argmin\limits_{\bm{v}\in\mathcal{V}}D(p_{\bm{X}}^*(\cdot), h(\cdot;\bm{v})) \notag \\
&= \argmin\limits_{\bm{v}\in\mathcal{V}}-\mathbb{E}_{p_{\bm{X}}^*}[\ln(h(\bm{X}; \bm{v}))] \notag \\ 
\label{Eq: CE optimization prob}
&= \argmax\limits_{\bm{v}\in\mathcal{V}} 
\mathbb{E}_{p_{\bm{X}}}[ \mathbb{I} \{ g(\bm{X})\leq0 \} \ln(h(\bm{X}; \bm{v})) ].
\end{align} 
The problem in Eq.~(\ref{Eq: CE optimization prob}) cannot be solved in closed form due to the indicator function inside the expectation, so instead we estimate $\bm{v}^*$ through optimizing an alternative objective function that substitutes the expectation in Eq.~(\ref{Eq: CE optimization prob}) with an IS estimator. 
That is, we solve 
\begin{equation}
\label{Eq: samp. counterpart of the CE optimization prob.}
\widehat{\bm{v}} = \argmax\limits_{\bm{v}\in\mathcal{V}} \frac{1}{N} \sum \limits_{i=1}^{N} \frac{p_{\bm{X}}(\bm{x}_i)\mathbb{I}\{g(\bm{x}_i)\leq 0\}}{p_{ ref}(\bm{x}_i)}\ln(h(\bm{x}_i; \bm{v})), \quad\quad \bm{x}_i \sim p_{ ref}(\cdot).
\end{equation}
$\{\bm{x}_i\}_{i=1}^N$ are samples from $p_{ ref}(\cdot)$, the IS distribution for estimating the expectation in Eq.~(\ref{Eq: CE optimization prob}), which is also known as the reference distribution \cite{DeBoer&others2005}. 
Note that $\widehat{\bm{v}}$ can be interpreted as the weighted MLE of the parametric family with weights $\{w_i \propto \frac{p_{\bm{X}}(\bm{x}_i)\mathbb{I}\{g(\bm{x}_i)\leq 0\}}{p_{ ref}(\bm{x}_i)}\}_{i=1}^N$ \cite{Geyer&others2019, Chan&others2023}.

As discussed in \cite{Chan&Kroese2012, Chan&others2023}, one should distinguish the sub-optimal IS distribution $h(\cdot;\bm{v}^*)$ from the chosen IS distribution $h(\cdot;\widehat{\bm{v}})$ in the CE method. 
$h(\cdot;\bm{v}^*)$ is conditional on the predefined parametric family while $h(\cdot;\widehat{\bm{v}})$ additionally depends on the CE procedure, in particular, the choice of the reference distribution $p_{ ref}(\cdot)$ and the number of samples. 
An appropriate reference distribution leads to an IS distribution $h(\bm{x}; \widehat{\bm{v}})$ close to $h(\bm{x};\bm{v}^*)$, which is the optimal choice within the given parametric family.
 
For rare event estimation, the reference distribution is chosen in an adaptive way. 
Let $p^{(t)}_{\bm{X}}(\cdot), t=1,...,T$ denote a sequence of intermediate target distributions that gradually approach the optimal IS distribution $p^*_{\bm{X}}(\cdot)$. 
The CE optimization problem is then solved iteratively for finding a good approximation to each $t$-th $p^{(t)}_{\bm{X}}(\cdot)$, and this results in a sequence of CE parameter vectors $\{\widehat{\bm{v}}^{(t)}, t=1,...,T \}$ and distributions $\{h(\cdot; \widehat{\bm{v}}^{(t)}), t=1,...,T \}$. 
The distribution we obtain in the $t$-th iteration, i.e., $h(\cdot; \widehat{\bm{v}}^{(t)})$, is used as the reference distribution $p_{ ref}(\cdot)$ for the CE procedure in iteration $t+1$. 
In this way, one takes $h(\cdot; \widehat{\bm{v}}^{(T-1)})$ as the reference distribution for Eq.~(\ref{Eq: samp. counterpart of the CE optimization prob.}), and $h(\cdot; \widehat{\bm{v}}^{(T)})$ as the final IS distribution. 
For the first iteration, the input distribution $p_{\bm{X}}(\cdot)$ is used as the reference distribution. 

There are many different ways of designing the intermediate target distributions \cite{DeBoer&others2005, Papaioannou&others2019, Uribe&others2021}. 
For instance, in the iCE method \cite{Papaioannou&others2019}, the intermediate target distribution reads
\begin{equation}
\label{Eq: intermediate target distribution of iCE method}
p^{(t)}_{\bm{X}}(\bm{x}) \triangleq \frac{1}{Z^{(t)}} p_{\bm{X}}(\bm{x})\Phi \left( -\frac{g(\bm{x})}{\sigma^{(t)}} \right), t=1,...,T
\end{equation}    
where $Z^{(t)}$ is the normalizing constant and $\Phi$ is the cumulative distribution function (CDF) of the standard normal distribution. 
The distribution sequence is driven by the parameter $\sigma^{(t)}>0$, and gradually approaches the optimal IS distribution with decreasing $\sigma^{(t)}$. 
The CE optimization problem for Eq.~(\ref{Eq: intermediate target distribution of iCE method}) reads  
\begin{equation}
\label{Eq: CE optimization for intermediate target dist. in iCE}
\bm{v}^{(t,*)} = \argmax\limits_{\bm{v}\in\mathcal{V}} 
\mathbb{E}_{p_{\bm{X}}}[ \Phi(-g(\bm{X})/\sigma^{(t)}) \ln(h(\bm{X}; \bm{v})) ],
\end{equation}
and the sample counterpart of Eq.~(\ref{Eq: CE optimization for intermediate target dist. in iCE}) can be written as 
\begin{align}
\label{Eq: samp. counterpart for intermediate target dist. in iCE}
\widehat{\bm{v}}^{(t)} = \argmax\limits_{\bm{v}\in\mathcal{V}} \frac{1}{N} & \sum \limits_{i=1}^{N} W(\bm{x}_i) \ln(h(\bm{x}_i; \bm{v})), \bm{x}_i \sim h(\cdot; \widehat{\bm{v}}^{(t-1)})\\
\label{Eq: weights in iCE}
W(\bm{x}_i) \triangleq & \frac{ p_{\bm{X}}(\bm{x}_i)\Phi(-g(\bm{x}_i)/\sigma^{(t)}) }{h(\bm{x}_i; \widehat{\bm{v}}^{(t-1)})}.
\end{align}
Note that $\widehat{\bm{v}}^{(t)}$ is the weighted maximum likelihood estimation (MLE) of $ \bm{v}^{(t,*)}$, and for a properly reparameterized exponential family, $\widehat{\bm{v}}^{(t)}$ is also the self-normalized IS estimator of $\bm{v}^{(t,*)}$ \cite{Chan&others2023}. 
The accuracy of $\widehat{\bm{v}}^{(t)}$ can be measured by the effective sample size (ESS), which is defined as the equivalent sample size required by MCS with the current target distribution to achieve the same variance as the self-normalized IS. 
The ESS of $\widehat{\bm{v}}^{(t)}$ in Eq.~(\ref{Eq: samp. counterpart for intermediate target dist. in iCE}) can be approximated by \cite{Kong1992}
\begin{equation}
\label{Eq: ESS}
ESS \approx \frac{N}{1+\widehat{\delta}^2 ( \{ W (\bm{x}_i)\}_{i=1}^N )}, \quad  \bm{x}_i \sim h(\cdot; \widehat{\bm{v}}^{(t-1)})
\end{equation}
where $\widehat{\delta}( \{ W(\bm{x}_i)\}_{i=1}^N )$ represents the sample coefficient of variation (c.o.v.) of the weights vector $\{ W(\bm{x}_i)\}_{i=1}^N$. 
Although the categorical mixture employed in this paper does not belong to the exponential family, we still expect that a large ESS will generally lead to a more accurate $\widehat{\bm{v}}^{(t)}$.
 
Given the reference distribution $h(\bm{x}_i; \widehat{\bm{v}}^{(t-1)})$, the iCE method fixes $N$ and changes $\sigma^{(t)}$ for achieving a constant ESS, and hence an accurate $\widehat{\bm{v}}^{(t)}$. 
Specifically, the intermediate target distribution in the iCE method is adapted at each $t$-th iteration by solving  
\begin{equation}
\label{Eq: the updating rule of sigma}
\sigma^{(t)} = \argmin \limits _{\sigma \in(0, \sigma^{(t-1)})} |\widehat{\delta} \left( \{ W(\bm{x}_i;\sigma) \}_{i=1}^{N} \right) -\delta_{tar}|, \quad\quad \bm{x}_i \sim h(\cdot; \widehat{\bm{v}}^{(t-1)}),
\end{equation}
where $\widehat{\delta}(\cdot)$ represents the sample c.o.v. of a vector and $\delta_{tar}$ is the hyperparameter that influences the convergence rate of the intermediate target distributions. 
A common choice is $\delta_{tar}=1.5$. 
The above procedure is iterated until 
\begin{equation}
\label{Eq: stopping criterion of iCE}
\widehat{\delta} \left( \left\lbrace \frac{\mathbb{I}\{g(\bm{x}_i)\leq0\}}{\Phi(-g(\bm{x}_i)/\sigma^{(t)})} \right\rbrace _{i=1}^N \right) \leq \delta_{\epsilon}, \quad\quad\bm{x}_i\sim h(\cdot; \widehat{\bm{v}}^{(t)}). 
\end{equation}
where $\delta_{\epsilon}$ is another hyperparameter and is often chosen to be the same as $\delta_{tar}$ \cite{Papaioannou&others2019}.

It should be stressed that the standard iCE method may suffer from overfitting when the sample size is small. 
To mitigate this issue, the BiCE method \cite{Chan&others2023} substitutes the weighted MLE with its Bayesian counterpart; therein the posterior predictive distribution is employed to update a single categorical parametric model in the context of network reliability assessment. 
In addition, the BiCE method employs an alternative weight function for solving $\sigma^{(t)}$ through Eq.~(\ref{Eq: the updating rule of sigma}), which is defined as 
\begin{equation}
\label{Eq: weights in BiCE}
W^{(alt)}(\bm{x}; \sigma) \triangleq \frac{\Phi(-g(\bm{x})/\sigma)}{\Phi(-g(\bm{x})/\sigma^{(t-1)} )}.
\end{equation} 
For a more detailed discussion and theoretical justification of Eq.~(\ref{Eq: weights in BiCE}), we refer to \cite{Chan&others2022b} and \cite{Chan&others2023}.
 
In this paper, we consider a more flexible parametric model, the categorical mixture, in the BiCE method. 
Before introducing the proposed CE approach, we first give an introduction to the categorical mixture model and its associated inference techniques in the following section.

\section{The categorical mixture model}
\label{sec: CatMix}
The categorical mixture model can be defined as:
\begin{equation}
\label{Eq: analytical expression of the mixture}
h_{cm}(\bm{x};\bm{\eta}) = \sum_{k=1}^{K}\alpha_k h_c(\bm{x};\bm{\theta}_k) = \sum_{k=1}^{K}\alpha_k \prod_{d=1}^D \prod_{j=1}^{n_d} \theta_{k,d,j}^{\mathbb{I}\{ x_d = s_{d,j}\}}.                    
\end{equation}
The probability distribution $h_{cm}(\cdot; \bm{\eta})$ is modelled as a linear combination of $K$ independent categorical components, denoted here as $h_c(\cdot; \bm{\theta})$. 
In this paper, $h_c(\cdot; \bm{\theta})$ denotes the independent categorical distribution with parameters $\bm{\theta}$. 
Specifically, in the $k$-th mixture component, the probability that the $d$-th component $X_d$ takes the $j$-th state $s_{d,j}$ is $\theta_{k,d,j}$, where $k=1,...,K; d=1,...,D; j=1,...,n_d$. 
$D$ and $n_d$ denote the number of input random variables $X_d$ and the number of states for each $X_d$. 
$\alpha_k, k=1,...K$, are the non-negative mixture weights that sum to one. 
All model parameters are collected in the vector $\bm{\eta}$, i.e., $\bm{\eta} \triangleq \{\alpha_k, \bm{\theta_k})\}_{k=1}^K$.
  
The mixture model described in Eq.~(\ref{Eq: analytical expression of the mixture}) is invariant with respect to the permutation of the component labels. 
As a result, the parameter estimation is unidentifiable \cite{Fruhwirth&others2019}. 
Additionally, Eq.~(\ref{Eq: analytical expression of the mixture}) remains invariant also (1) when adding a mixture component with zero weight, or (2) when replicating any of the mixture components and splitting the associated weight \cite{Fruhwirth&others2019}, which leads to a broader class of unidentifiability of the model parameters \cite{Rousseau&Mengersen2011}. 

\subsection{MLE of the categorical mixture and EM algorithm}
\label{Subsec: MLE and EM}
Suppose we want to fit a categorical mixture described in Eq.~(\ref{Eq: analytical expression of the mixture}) with $N$ samples, $\mathcal{X}\triangleq\{\bm{x}_i\}_{i=1}^N$, and consider the case where the number of mixture components is known to be $K$. 
The most common approach is through MLE. 
The log-likelihood is
\begin{equation}
\label{Eq: LL of CatMix}
\ln\mathcal{L}(\bm{\eta};\mathcal{X}) \triangleq \ln \left( \prod_{i=1}^N h_{cm}(\bm{x}_i;\bm{\eta}) \right) = \sum_{i=1}^N \ln \left( \sum_{k=1}^{K}\alpha_k h_c(\bm{x}_i;\bm{\theta}_k) \right).
\end{equation} 
The MLE for the categorical mixture cannot be obtained in closed form. 
If one observes the allocation variable $z_i$ for each $i$-th sample $\bm{x}_i$, the log-likelihood function in Eq.~(\ref{Eq: LL of CatMix}) takes the following form:  
\begin{equation}
\label{Eq: LL of CatMix with complete data}
\ln\mathcal{L}^{(c)}(\bm{\eta};\mathcal{X}) = \sum_{i=1}^N \ln \left( \alpha_{z_i} h_c(\bm{x}_i;\bm{\theta}_{z_i}) \right) = \sum_{k=1}^{K} \sum_{i \in \mathcal{C}_k} \ln \left( \alpha_k h_c(\bm{x}_i;\bm{\theta}_k) \right).
\end{equation}
The allocation variable $z_i$ specifies which mixture component generates $\bm{x}_i$, and $\mathcal{C}_k \triangleq \{i: i=1,...,N, z_i=k\}$ collects the indexes of all the samples generated by the $k$-th component of the mixture. 
Eq.~(\ref{Eq: LL of CatMix with complete data}) is often termed the complete data log-likelihood in the context of MLE to differentiate it from the log-likelihood in Eq.~(\ref{Eq: LL of CatMix}). 
Maximizing Eq.~(\ref{Eq: LL of CatMix with complete data}), is equivalent to fitting a categorical distribution $h_c(\cdot;\bm{\theta}_k)$ for each $\mathcal{C}_k$ and letting the associated weight $\alpha_k$ be proportional to $|\mathcal{C}_k|$, the number of samples in $\mathcal{C}_k$.
Note that the closed-form solution to the MLE is well known for the single categorical distribution.
 
However, the allocation variables $\{z_i\}_{i=1}^N$ are not observed; they are latent variables. 
One approach is to estimate the latent variables through a clustering algorithm. 
However, clustering of categorical data is usually not straightforward, especially in the high-dimensional sample space.
 
For finding a mode of the log-likelihood function shown in Eq.~(\ref{Eq: LL of CatMix}), one usually resorts to the EM algorithm, which iteratively updates and optimizes the so-called $Q$ function, an auxiliary function that computes the expectation of the complete data log-likelihood in Eq.~(\ref{Eq: LL of CatMix with complete data}). That is, 
\begin{align}
\label{Eq: Q function for getting the MLE}
Q( \bm{\eta} ; \{ p_{Z_i}(\cdot) \}_{i=1}^N ) 
&= \sum_{i=1}^N  \mathbb{E}_{Z_i \sim p_{Z_i}(\cdot)} \left[ \ln \left( \alpha_{Z_i} h_c(\bm{x}_i; \bm{\theta}_{Z_i}) \right) \right] \notag \\
&= \sum_{i=1}^N  \sum_{k=1}^K p_{Z_i}(k) \ln \left( \alpha_k h_c(\bm{x}_i; \bm{\theta}_k) \right),
\end{align}
where $p_{Z_i}(\cdot)$ is a customary distribution for the $i$-th allocation variable $Z_i$. 
$p_{Z_i}(k)$ represents the probability that the $i$-th sample is generated by the $k$-th component of the mixture.
Note that $p_{Z_i}(\cdot)$ can be an arbitrary distribution without necessarily being related to $\bm{\eta}$.
According to Jensen's inequality, the log-likelihood function $\ln\mathcal{L}(\bm{\eta}; \mathcal{X})$ in Eq.~(\ref{Eq: LL of CatMix}) is bounded from below by the $Q$ function plus a constant \cite{Murphy2012}. 
That is
\begin{align}
\label{InEq: lower bound of the LL}
\ln\mathcal{L}(\bm{\eta};\mathcal{X}) 
&= \sum_{i=1}^N \ln \left( \sum_{k=1}^K p_{Z_i}(k) \frac{ \alpha_k h_c(\bm{x}_i; \bm{\theta}_k) }{p_{Z_i}(k)}\right) \notag \\ 
&\geqslant \sum_{i=1}^N \left( \sum_{k=1}^K p_{Z_i}(k) \ln \frac{ \alpha_k h_c(\bm{x}_i; \bm{\theta}_k) }{p_{Z_i}(k)}\right) \notag \\
&= Q( \bm{\eta} ; \{ p_{Z_i}(\cdot) \}_{i=1}^N ) + \sum_{i=1}^N \mathbb{H}(p_{Z_i}(\cdot)).
\end{align} 
$\mathbb{H}(p_{Z_i}(\cdot)) \triangleq \sum_{k=1}^K -p _{Z_i}(k) \ln (p_{Z_i}(k)) \geqslant 0$ is the entropy of the distribution $p_{Z_i}(\cdot)$ and is a constant with respect to $\bm{\eta}$. 
The inequality (\ref{InEq: lower bound of the LL}) takes the equal sign if 
\begin{equation}
\label{Eq: optimal distribution of Z_i}
p_{Z_i}(k) = \frac{ \alpha_k h_c(\bm{x}_i; \bm{\theta}_k) }{\sum_{k'=1}^N \alpha_{k'} h_c(\bm{x}_i; \bm{\theta}_{k'})} \triangleq \gamma_{i,k}(\bm{\eta})
\end{equation}
holds for each $k=1,...,K$ and $i=1,...,N$. $[\gamma_{i,k}(\bm{\eta})]_{N \times K}$ is also termed the responsibility matrix in the literature \cite{Murphy2012}.
 
Eq.~(\ref{Eq: optimal distribution of Z_i}) indicates that, for any given $\bm{\eta}$ denoted as $\bm{\eta}^{(cur)}$, one can choose $p_{Z_i}(\cdot)=\gamma_{i,\cdot}(\bm{\eta}^{(cur)})$ for each $Z_i$, such that $\ln\mathcal{L}(\bm{\eta}^{(cur)}; \mathcal{X}) = Q\left( \bm{\eta}^{(cur)}; \bm{\eta}^{(cur)} \right) + C(\bm{\eta}^{(cur)})$, where $Q\left( \bm{\eta}^{(cur)}; \bm{\eta}^{(cur)} \right)$ is short for $Q\left( \bm{\eta}^{(cur)}; \{ \gamma_{i,\cdot}(\bm{\eta}^{(cur)}) \}_{i=1}^N \right) $ and $C(\bm{\eta}^{(cur)}) \triangleq \sum_{i=1}^N \mathbb{H}(\gamma_{i,\cdot}(\bm{\eta}^{(cur)}))$. 
This is also known as the expectation step (E step) of the EM algorithm, in which we compute the responsibility matrix $[\gamma_{i,k}(\bm{\eta}^{(cur)})]_{N \times K}$ via Eq.~(\ref{Eq: optimal distribution of Z_i}) and formulate the $Q$ function.

In the next step, the maximization step or the M step for short, the EM algorithm fixes $p_{Z_i}(k) = \gamma_{i,k}(\bm{\eta}^{(cur)})$ for each $i$ and $k$ and maximizes the $Q$ function over $\bm{\eta}$ to find a new $\bm{\eta}$ denoted as $\bm{\eta}^{(nxt)}$ whose $Q$ function is larger than that of $\bm{\eta}^{(cur)}$, i.e., $Q\left( \bm{\eta}^{(nxt)}; \bm{\eta}^{(cur)} \right)\geqslant Q\left( \bm{\eta}^{(cur)}; \bm{\eta}^{(cur)} \right)$. 
Since the $Q$ function (plus a constant) is a lower bound of the log-likelihood as shown in Inequality(\ref{InEq: lower bound of the LL}), the log-likelihood of $\bm{\eta}^{(nxt)}$ is also larger than that of $\bm{\eta}^{(cur)}$. In fact, we have $\ln\mathcal{L}(\bm{\eta}^{(nxt)}; \mathcal{X}) \geqslant Q\left( \bm{\eta}^{(nxt)}; \bm{\eta}^{(cur)} \right) + C(\bm{\eta}^{(cur)}) \geqslant Q\left( \bm{\eta}^{(cur)}; \bm{\eta}^{(cur)} \right) + C(\bm{\eta}^{(cur)})= \ln\mathcal{L}(\bm{\eta}^{(cur)}; \mathcal{X}) $.
The point here is that optimizing the $Q$ function is much easier than optimizing the log-likelihood function in Eq.~(\ref{Eq: LL of CatMix}). 
Specifically, the M step solves the following optimization problem:
\begin{align}
\label{Eq: optimization problem of maximizing Q function}
\bm{\eta}^{(nxt)}
&= \argmax_{\bm{\eta}} Q(\bm{\eta}; \bm{\eta}^{(cur)} ) \notag \\ 
&= \argmax_{\bm{\eta}} \sum_{i=1}^N  \sum_{k=1}^K \gamma_{i,k}(\bm{\eta}^{(cur)}) \ln \left( \alpha_k h_c(\bm{x}_i; \bm{\theta}_k) \right).  
\end{align}
For the categorical mixture shown in Eq.~(\ref{Eq: analytical expression of the mixture}), the closed-form solution $\bm{\eta}^{(nxt)}=\{ \alpha^{(nxt)}_k, \bm{\theta}^{(nxt)}_k \}_{k=1}^K$ to the optimization problem in Eq.~(\ref{Eq: optimization problem of maximizing Q function}) exists and is given by:
\begin{align}
\label{Eq: updating rule of the alpha}
\alpha^{(nxt)}_k &= \frac{\sum_{i=1}^N  \gamma_{i,k}(\bm{\eta}^{(cur)})}{ \sum_{k=1}^K \sum_{i=1}^N \gamma_{i,k}(\bm{\eta}^{(cur)}) }, \\ 
\label{Eq: updating rule of the theta}
\theta^{(nxt)}_{k,d,j} &= \frac{\sum_{i=1}^N  \gamma_{i,k}(\bm{\eta}^{(cur)}) \mathbb{I}\{ x_{i,d}=s_{d,j} \}}{\sum_{i=1}^N \gamma_{i,k}(\bm{\eta}^{(cur)}) }.  
\end{align}
Note that if there is no sample equal to $s_{d,j}$, the probability assigned to that state, i.e., $\theta^{(t+1)}_{k,d,j}$, will become zero in each $k$-th mixture component, and this can lead to overfitting, as will be shown later in Sec.~4.1.

Through iterating the above two steps by setting $\bm{\eta}^{(cur)} = \bm{\eta}^{(nxt)}$, one ends up with a sequence of model parameters, $\bm{\eta}^{(0)}, \bm{\eta}^{(1)},...,\bm{\eta}^{(T)} $, that gradually improves the log-likelihood function. 
Although this does not strictly imply the convergence of the EM algorithm to a local maximum, usually this is the case.

$\bm{\eta}^{(0)}$ represents an initial guess of the model parameters. 
Given the sample set and the stopping criteria, the final estimate of the model parameters only relates to the choice of $\bm{\eta}^{(0)}$.
A common strategy for getting an appropriate starting point is to first launch several short pilot runs of the EM algorithm, each with a different initialization, and then to choose the starting point for which the log-likelihood is the largest. 
It is noted that the EM algorithm can also start from the M step instead of the E step, which requires an initial guess of the $p_{Z_i}(\cdot)$ for each $Z_i$. 


\subsection{Bayesian inference}
\label{Subsec: Bayesian inference}
In the following, we adopt the Bayesian viewpoint to the inference of mixture models with $K$ components and interpret the model parameters as random variables, $\bm{E}$, whose prior distribution is denoted as $p_{\bm{E}}(\bm{\eta})$. The posterior distribution of parameters $\bm{E}$ given $\mathcal{X}$ is given by Bayes' rule as 
\begin{equation}
\label{Eq: posterior distribution}
p_{\bm{E}|\mathcal{X}}(\bm{\eta}|\mathcal{X}) = \frac{ \mathcal{L}(\bm{\eta}|\mathcal{X}) \cdot p_{\bm{E}}(\bm{\eta})}{p_{\mathcal{X}}(\mathcal{X})}.
\end{equation}
The resulting predictive distribution reads
\begin{equation}
\label{Eq: predictive distribution}
p_{\bm{X}|\mathcal{X}}(\bm{x}|\mathcal{X}) = \int_{\Omega_{\bm{E}}}h_{cm}(\bm{x}|\bm{\eta}) \cdot p_{\bm{E}|\mathcal{X}}(\bm{\eta}|\mathcal{X}) d\bm{\eta},
\end{equation}
which is an expectation of the mixture model with respect to the posterior distribution of model parameters. $\Omega_{\bm{E}}$ represents the sample space of $\bm{E}$. 
The posterior distribution, and hence also the predictive distribution, is not analytically tractable. Instead, the posterior distribution can be approximated through MCMC sampling, 
\begin{equation}
\label{Eq: sample approximation of the posterior}
p_{\bm{E}|\mathcal{X}}(\bm{\eta}|\mathcal{X}) \approx \frac{1}{N_p} \sum_{i=1}^{N_p} \delta (\bm{\eta} - \bm{\eta}_i ),
\end{equation}
where $\delta(\cdot)$ is the Dirac delta function and $\{\bm{\eta}_i\}_{i=1}^{N_p}$ denotes the posterior samples.      
In this way, the predictive distribution is a mixture of mixtures consisting of a total of $N_p \cdot K$ mixture components. 
The computational cost of computing and sampling from this approximate predictive distribution is roughly $N_p$ times the cost for a $K$-component mixture, and $N_p$ is often large, say thousands.   
Therefore in this paper, we resort to a single point estimate of the model parameter, namely the MAP estimate $\widetilde{\bm{\eta}}$, for which the posterior distribution $p_{\bm{E}|\mathcal{X}}(\bm{\eta}|\mathcal{X})$ is maximized.
Another benefit of using the MAP is that it can be obtained directly from the EM algorithm \cite{Murphy2012}, which is significantly cheaper than running an MCMC algorithm.
 
The derivative of the EM algorithm for computing the MAP estimate follows the same lines as for the MLE, with a minor modification to account for the prior. 
Specifically, a log-prior distribution $\ln(p_{\bm{E}}(\bm{\eta}))$ is added to the original $Q$ function in Eq.~(\ref{Eq: Q function for getting the MLE}), and the EM algorithm proceeds iteratively with the following two steps:  
(1) E step: compute the distribution of the allocation variables $\bm{Z}$ through Eq.~(\ref{Eq: optimal distribution of Z_i}).
(2) M step: update the model parameters through maximizing a modified $Q$ function, i.e., 
\begin{align}
\label{Eq: Q function for getting the MAP}
\bm{\eta}^{(nxt)} = \argmax_{\bm{\eta}} \sum_{i=1}^N  \sum_{k=1}^K \gamma_{i,k}(\bm{\eta}^{(cur)}) \ln \left( \alpha_k h_c(\bm{x}_i|\bm{\theta}_k) \right) + \ln(p_{\bm{E}}(\bm{\eta})). 
\end{align}
In particular, for a conjugate prior distribution $p_{\bm{E}}(\bm{\eta})$, a closed-form updating scheme can be derived for the categorical mixture parameters. 

\subsection{Model selection and BIC}
\label{Subsec: Model selection and BIC}
In this subsection, we discuss how to select the number of components $K$ in the mixture model $h_{cm}(\cdot; \bm{\eta})$ using the information provided by the samples $\mathcal{X} \triangleq \{\bm{x}_i\}_{i=1}^N$. 
Let the initial pool of candidate models be $\{ \mathcal{M}_K \}_{k=1}^{K_{max}}$ where $\mathcal{M}_K$ refers to a mixture of $K$ independent categorical components and $K_{max}$ is a hyperparameter representing the maximum number of mixture components. 
From a Bayesian perspective, we favor the model $\mathcal{M}_{\widetilde{K}}$ with the highest posterior probability, or equivalently with the highest log-posterior. That is 
\begin{align}
\label{Eq: MAP of the K}
\widetilde{K}
&= \argmax_K \hspace{0.3em} \ln p_{\mathcal{M}|\mathcal{X}}(\mathcal{M}_K|\mathcal{X}) \notag \\ 
&= \argmax_K \hspace{0.3em} \ln \mathcal{L}(\mathcal{M}_K|\mathcal{X}) + \ln p_{\mathcal{M}}(\mathcal{M}_K) \notag \\
&= \argmax_K \hspace{0.3em} \ln \left( \int_{\Omega_{\bm{E}}} \mathcal{L}(\bm{\eta}|\mathcal{X}, \mathcal{M}_K) p_{\bm{E}|\mathcal{M}}(\bm{\eta}|\mathcal{M}_K) d\bm{\eta} \right) + \ln p_{\mathcal{M}}(\mathcal{M}_K).  
\end{align}
Here, $p_{\mathcal{M}}(\mathcal{M}_K)$ represents the prior probability for each $k$-th candidate model, and it is often assumed to be uniformly distributed among all candidates. 
$\mathcal{L}(\mathcal{M}_K|\mathcal{X})$ denotes the integrated likelihood, or the marginal likelihood, and is the integral of the likelihood function $\mathcal{L}(\bm{\eta}|\mathcal{X}, \mathcal{M}_K)$ multiplied by the parameter prior distribution $p_{\bm{E}|\mathcal{M}}(\bm{\eta}|\mathcal{M}_K)$ over the whole sample space of the parameters $\Omega_{\bm{E}}$. Note that this is actually the normalizing constant of the posterior distribution of the parameters in $\mathcal{M}_K$, i.e., $p_{\bm{E}|\mathcal{X},\mathcal{M}}(\bm{\eta}|\mathcal{X}, \mathcal{M}_K)$.
 
Computing the integrated likelihood involves a high dimensional integration whose closed-form solution is not available. 
Nevertheless, it can be approximated through various sampling-based methods \cite{Gelfand&Dey1994, Fruhwirth2004, meng&Wong1996}. 
These methods often rely on computationally expensive MCMC algorithms and are limited to a small $K$, for example, up to 6 \cite{Fruhwirth&others2019}. 
The Bayesian information criterion (BIC) serves as a crude but computationally cheap proxy of the log-posterior probability when $p_{\mathcal{M}}(\mathcal{M}_K)\propto 1$. 
BIC was first introduced by Schwarz \cite{Schwarz1978} for asymptotically approximating the log-posterior probability of a linear model given observations $\mathcal{X}$ from a regular exponential family (see the definition in \cite{Schwarz1978}); 
therein the BIC is defined as $\ln\mathcal{L}(\widehat{\bm{\eta}}|\mathcal{X}, \mathcal{M})) - \frac{\text{dim}(\mathcal{M})\ln(N)}{2}$, where $\ln\mathcal{L}(\widehat{\bm{\eta}}|\mathcal{X}, \mathcal{M})$ represents the mode of the log-likelihood function evaluated at the MLE point $\widehat{\bm{\eta}}$, and $\text{dim}(\mathcal{M})$ denotes the number of free parameters in $\mathcal{M}$. 
Another commonly used definition is given by 
\begin{equation}
\label{Eq: BIC}
\text{BIC}(\mathcal{M}) \triangleq -2\ln\mathcal{L}(\widehat{\bm{\eta}}|\mathcal{X}, \mathcal{M}) + \text{dim}(\mathcal{M})\ln(N).
\end{equation}
Note that under the definition of Eq.~(\ref{Eq: BIC}), the model with the smallest BIC is favored.

The derivation of the BIC relies on the Laplace approximation to the likelihood function $ \mathcal{L}(\bm{\eta}|\mathcal{X}, \mathcal{M})$, which does not apply to multi-modal posterior distributions, and thus BIC cannot be interpreted as a meaningful approximation to the log-posterior of a mixture model. 
In spite of this, BIC remains one of the state-of-art techniques for selecting the number of mixture components in practice\cite{Roeder&Wasserman1997, Steele&Raftery2010, Baudry&Celeux2015, Fruhwirth&others2019}. 
Additionally, BIC can be computed directly as a by-product of the EM algorithm without employing any computationally expensive MCMC algorithm. 
Therefore, throughout this paper, we adopt the BIC as the model selection technique.
\section{Bayesian improved cross entropy method with the categorical mixture model}
\label{sec: BiCE with CatMix}
In this section, we introduce the Bayesian iCE method with the categorical mixture model for network reliability analysis. 
With slight abuse of notation, we omit the subscript for all prior and posterior distributions, and use, e.g., $p(\bm{\eta})$ to represent $p_{\bm{E}}(\bm{\eta})$.
\subsection{Motivation}
\label{Subsec: motivation}
As mentioned in Sec.~\ref{sec: CE based IS}, the ‘distance’ between the optimal IS distribution and the suboptimal IS distribution is only related to the chosen parametric model. 
For a fixed parametric model, the ‘distance’ remains fixed assuming that the CE optimization problem is solved exactly. 
An inappropriate parametric model will lead to an IS estimator with large variance in the final level of CE-based methods. 
In particular, this can happen when approximating an optimal IS distribution that implies a strong dependence between component states with the independent categorical model.

To account for the dependence between the component states, one could use a dependent categorical distribution. 
However, it is not straightforward to choose an appropriate dependence structure that is both easy to sample from and convenient to update. 
Instead, we consider the mixture of independent categorical distributions as the parametric model. 
The flexibility of this mixture model enables capturing arbitrary dependencies between variables in the optimal IS distribution. 
In the CE-based IS, the parametric model is updated by maximizing a weighted log-likelihood function as shown in Eq.~(\ref{Eq: samp. counterpart for intermediate target dist. in iCE}). 
Therefore, techniques for MLE can also be used in the CE-based methods with minor modifications to account for the weights. 
For instance, Geyer et.al., \cite{Geyer&others2019} used the EM algorithm for updating a Gaussian mixture model in the CE method. They found that the Gaussian mixture model performs consistently worse than the single Gaussian model especially when the sample size is small. 
The reason is that the EM algorithm tends to overfit the weighted samples and hence it is more sensitive to sample sets that misrepresent the target distribution.  
This can happen when the geometry/shape of the intermediate target distributions changes significantly in CE-based methods, which results in one or more modes of the target distribution being missing or cannot be sufficiently reflected by the weighted samples. 
The overfitting issue is even more severe for updating the categorical mixture in CE methods.
If there is no sample falling into a certain category during the adaptive process, the probability assigned to this category will be zero for all mixture components, resulting in a potentially biased estimate of the final IS estimator. 
This is also known as the zero count problem in the context of MLE with categorical data \cite{Murphy2012}. A detailed discussion of the zero count problem for the CE method with the independent categorical parametric model can be found in \cite{Chan&others2023}. 

\subsection{Bayesian updating for cross-entropy-based methods}
\label{Subsec: MAP for CatMix in iCE}
\subsubsection{The basic idea}
To circumvent the overfitting issue of the weighted MLE, we propose to use the Bayesian approach for updating the categorical mixture in the CE method. 
At each level, we approximate the weighted MAP of a $K$-component mixture, denoted as $\widetilde{\bm{\eta}}|\mathcal{M}_K$, through a generalized version of the EM algorithm that works with weighted samples. 
Here, we use 'approximate' to indicate that the algorithm is prone to get stuck in a local maximum, but this limitation can be alleviated by launching short pilot runs as mentioned in Subsec.~\ref{Subsec: MLE and EM}. 
Model selection is performed for estimating the optimal number of components $\widetilde{K}$ in the categorical mixture, whereby the number of mixture components leading to the smallest BIC is selected. 
Next, we employ the $\widetilde{K}$-component categorical mixture with its parameters fixed at $\widetilde{\bm{\eta}}|\mathcal{M}_{\widetilde{K}}$ as the reference/sampling distribution at the $(t+1)$-th level in the CE method. 
We term the proposed method BiCE-CM. 
\subsubsection{The generalized EM algorithm}
In this subsection, we introduce a generalized version of the EM algorithm and demonstrate its properties.  
To this end, we first attach a Dirichlet prior, which is the conjugate prior for categorical distributions, to each model parameter, i.e., 
\begin{align}
\label{Eq: prior for catMix}
\bm{\alpha} \triangleq \{\alpha_k \} & _{k=1}^K \sim \text{Dir}(\cdot|\bm{a}) \notag \\
\bm{\theta}_{k,d} \triangleq \{ \theta_{k,d,j} & \}_{j=1}^{n_d} \sim \text{Dir} (\cdot|\bm{b}_{k,d}) \notag \\
p(\bm{\eta}|\mathcal{M}_K) = \text{Dir} & (\bm{\alpha}|\bm{a}) \prod_{k=1}^K \prod_{d=1}^D \text{Dir}(\bm{\theta}_{k,d}|\bm{b}_{k,d}) ,
\end{align}
where $\bm{a}=(a_1,...,a_k)$ and $\bm{b}_{k,d}=(b_{k,d,1},...,b_{k,d,n_d})$ are predefined concentration parameters. 
We obtain an MAP estimate of the model parameters $\bm{\eta}$ through maximizing the weighted log-posterior distribution $\ln \left( p^{(w)}(\bm{\eta}|\mathcal{X},\mathcal{M}_K) \right)$, which reads:
\begin{align}
\label{Eq: weighted log-posterior}
\ln \left( p^{(w)}(\bm{\eta}|\mathcal{X},\mathcal{M}_K) \right) 
&= \ln\mathcal{L}^{(w)}(\bm{\eta}|\mathcal{X},\mathcal{M}_K)) + \ln(p(\bm{\eta}|\mathcal{M}_K)) \notag\\
&=\sum_{i=1}^N w_i \ln \left( h_{cm}(\bm{x}_i|\bm{\eta}) \right) + \ln(p(\bm{\eta}|\mathcal{M}_K)),
\end{align}
where $\mathcal{L}^{(w)}(\bm{\eta}|\mathcal{X},\mathcal{M}_K)$ is the weighted likelihood with $w_i \triangleq \frac{ N W(\bm{x}_i)}{\sum_{j=1}^N W(\bm{x}_j)}$ representing the normalized weight of the $i$-th sample $\bm{x}_i$; herein, the weight function $W(\cdot)$ defined in Eq.~(\ref{Eq: weights in iCE}) is normalized such that the sum of the weights is equal to $N$. 
Note that normalizing the weights $\{ W(\bm{x}_i) \}_{i=1}^{N}$ does not change the solution to the original CE optimization problem in Eq. (\ref{Eq: samp. counterpart for intermediate target dist. in iCE}), i.e., $\widehat{\bm{v}}^{(t)}$, but can modify the relative strength between the log-prior and the weighted log-likelihood term in Eq.~(\ref{Eq: weighted log-posterior}). 
As the sample size $N$ increases, the log-prior term will be dominated by the weighted log-likelihood, and hence, the solution to Eq.~(\ref{Eq: weighted log-posterior}) coincides with the results obtained from Eq.~(\ref{Eq: samp. counterpart for intermediate target dist. in iCE}) in large sample settings. 
On the other hand, when the sample size is small/moderate, the prior term serves as a regularizer that penalizes the weighted log-likelihood. 
Different kinds of prior distributions or regularizers can be applied depending on the problems at hand, but a detailed investigation is left for future work. In this paper, we focus on the Dirichlet prior as shown in Eq.~(\ref{Eq: prior for catMix}).
 
A generalized version of the EM algorithm is employed to maximize Eq.~(\ref{Eq: weighted log-posterior}), which iteratively updates the following weighted $Q$ function
\begin{align}
\label{Eq: weighted Q function including the prior}
&Q^{(w)}(\bm{\eta}; \{ p_{Z_i}(\cdot) \}_{i=1}^N)  \triangleq \sum_{i=1}^N w_i \mathbb{E}_{Z_i \sim p_{Z_i}(\cdot)} \left[ \ln \left( \alpha_{Z_i} h_c(\bm{x}_i; \bm{\theta}_{Z_i}) \right) \right] + \ln(p(\bm{\eta}|\mathcal{M}_K)) \notag \\
& \quad \quad \quad \quad \quad \quad \quad = \sum_{i=1}^N w_i \sum_{k=1}^K p_{Z_i}(k) \left[ \ln \left( \alpha_k h_c(\bm{x}_i; \bm{\theta}_k \right) \right] + \ln(p(\bm{\eta}|\mathcal{M}_K)).
\end{align} 
In the E step, we compute the responsibility matrix $[\gamma_{i,k}(\bm{\eta}^{(cur)})]_{N\times K}$ via Eq.~(\ref{Eq: optimal distribution of Z_i}) and formulate $Q^{(w)}(\bm{\eta}; \bm{\eta}^{(cur)}) \triangleq Q^{(w)}(\bm{\eta}; \{ \gamma_{i, \cdot}(\bm{\eta}^{(cur)}) \}_{i=1}^N)$; in the M step, we maximize $Q^{(w)}(\bm{\eta}; \bm{\eta}^{(cur)})$ over $\bm{\eta}$, resulting in the following updating scheme for the categorical mixture:
\begin{align}
\label{Eq: weighted MAP for alpha}
\alpha^{(nxt)}_k 
&= \frac{\sum_{i=1}^N  w_i \gamma_{i, k}(\bm{\eta}^{(cur)})+a_k-1}{ \sum_{k=1}^K \sum_{i=1}^N w_i \gamma_{i, k}(\bm{\eta}^{(cur)}) + \sum_{k=1}^K a_k - K }, \\
\label{Eq: weighted MAP for theta}
\theta^{(nxt)}_{k,d,j} &= \frac{\sum_{i=1}^N  w_i \gamma_{i, k}(\bm{\eta}^{(cur)}) \mathbb{I}\{ x_{i,d}=s_{d,j} \} + b_{k,d, j}-1}{\sum_{i=1}^N w_i \gamma_{i, k}(\bm{\eta}^{(cur)}) + \sum_{j=1}^{n_d} b_{k,d,j}-n_d}. 
\end{align}
Similarly to the original EM algorithm, it holds that
\begin{align}
& \ln \left( p^{(w)}(\bm{\eta}^{(nxt)} |\mathcal{X}, \mathcal{M}_K) \right) 
\geqslant Q^{(w)}(\bm{\eta}^{(nxt)}; \bm{\eta}^{(cur)}) + C^{(w)}(\bm{\eta}^{(cur)})\notag \\
& \quad \quad \geqslant Q^{(w)}(\bm{\eta}^{(cur)}; \bm{\eta}^{(cur)})+ C^{(w)}(\bm{\eta}^{(cur)}) = \ln \left( p^{(w)}(\bm{\eta}^{(cur)}|\mathcal{X},\mathcal{M}_K) \right), 
\end{align}
where $C^{(w)}(\bm{\eta}^{(cur)}) \triangleq \sum_{i=1}^N w_i \mathbb{H}(\gamma_{i,\cdot}(\bm{\eta}^{(cur)})) $. 
We end up with a sequence of parameters $\bm{\eta}^{(0)},...,\bm{\eta}^{(T)}$ that converges to one of the modes (or saddle points) of the weighted log-posterior distribution, and $\bm{\eta}^{(T)}$ is regarded as an approximate weighted MAP, $\widetilde{\bm{\eta}}|\mathcal{M}_K$.
\subsubsection{The weighted MAP mitigates the overfitting and is unbiased}
$\bm{\eta}^{(T)}$ can be written as a linear combination of a data-dependent estimate $\bm{\eta}^{(T;\text{D})}$, which exploits the current data, and a user-defined prior estimate $\bm{\eta}^{(T;\text{pri})}$, which can be designed to explore a wider part of the sample space and thus is capable of finding potentially missing modes. 
Taking $\theta^{(T)}_{k,d,j}$ as an example, let $nxt=T, cur=T-1$ and rearrange Eq.~(\ref{Eq: weighted MAP for theta}) as follows:
\begin{equation}
\label{Eq: rephrasing of the updating formula of theta}
\theta^{(T)}_{k,d,j} = \lambda_{k,d}(\bm{\eta}^{(T-1)}) \theta_{k, d, j}^{(nxt;\text{D})} + (1-\lambda_{k,d}(\bm{\eta}^{(T-1)})) \theta_{k, d, j}^{(\text{pri})}.
\end{equation}
where $\theta_{k, d, j}^{(T;\text{D})} \triangleq \frac{\sum_{i=1}^N  w_i \gamma_{i, k}(\bm{\eta}^{(T-1)}) \mathbb{I}\{ x_{i,d}=s_{d,j} \}}{\sum_{i=1}^N w_i \gamma_{i, k}(\bm{\eta}^{(T-1)})}$, and $\theta_{k,d,j}^{(\text{pri})} \triangleq \frac{ b_{k,d,j} - 1 }{\sum_{j=1}^{n_d} b_{k,d,j}-n_d}$. 
$\theta_{k, d, j}^{(T;\text{D})}$ and $\theta_{k,d,j}^{(\text{pri})}$ are combined via 
\begin{equation}
\label{Eq: mixing factor}
\lambda_{k,d}(\bm{\eta}^{(T-1)}) \triangleq \frac{\sum_{i=1}^N  w_i \gamma_{i, k}(\bm{\eta}^{(T-1)})}{\sum_{i=1}^N  w_i \gamma_{i, k}(\bm{\eta}^{(T-1)})+\sum_{j=1}^{n_d} b_{k,d,j}-n_d},
\end{equation}
which is a factor indicating the relative strength of the data with respect to the combined information from the data and prior. $\lambda_{k,d}(\bm{\eta}^{(T-1)})$ tunes the exploitation and exploration behaviour of $\theta^{(T)}_{k,d,j}$; the larger $\lambda_{k,d}(\bm{\eta}^{(T-1)})$ is, the more dominant is $\theta_{k, d, j}^{(T;\text{D})}$ in Eq.~(\ref{Eq: rephrasing of the updating formula of theta}). 
A similar interpretation also applies to $\alpha_k^{(T)}$.  
Moreover, if we set $b_{k,d,j}>1$ for each $k$,$d$ and $j$, $\theta^{(T)}_{k,d,j}$ is always positive even when no samples fall into the category $s_{d, j}$, i.e., the zero count issue is mitigated in small sample settings. 
As a result, the sample space of the reference distribution at each intermediate level will no longer shrink even with a small number of samples, which ensures an \textbf{unbiased} IS estimator at the final CE level.
\subsubsection{Implementation details} 
\label{Subsubsec: Implementation details}
\paragraph{Initialization} 
To initialize the generalized EM algorithm, we launch several short pilot runs, each from a random realization of the responsibility matrix $[ \gamma_{i,k}^{(0)} ]_{N \times K}$. 
The $i$-th row of the responsibility matrix is a $K$-component vector generated uniformly and independently over the standard $(K-1)$-simplex, i.e., the vector follows the symmetric Dirichlet distribution $\text{Dir}(\cdot|[1,...,1])$.
The responsibility matrix that achieves the highest weighted log-posterior is chosen as the starting point from which we iteratively perform the M step and E step until convergence.
\paragraph{The prior distribution} 
For selecting an appropriate Dirichlet prior distribution in the BiCE-CM, we rearrange Eq.~(\ref{Eq: mixing factor}) as follows: 
\begin{equation}
\label{Eq: derivation of the prior distribution_step1}
\sum_{j=1}^{n_d} b_{k,d,j}-n_d = \left( 1-\lambda_{k,d}(\bm{\eta}^{T-1}) \right) \cdot \sum_{i=1}^N  w_i \gamma_{i, k} \left( \bm{\eta}^{T-1} \right).
\end{equation}
For simplicity, let $\gamma_{i, k}\left( \bm{\eta}^{T-1} \right) = 1/K$ for each $i$ and $k$, and assume a symmetric Dirichlet prior for each $\bm{\theta}_{k,d}$, i.e., $b_{k,d,j_1}=b_{k,d,j_2}$ for $1 \leqslant j_1 \neq j_2 \leqslant n_d$ and $1\leqslant k \leqslant K, 1 \leqslant d \leqslant D$. 
$\bm{\theta}_{k,d}$ represents the PMF of $X_d$ implied by the $k$-th component of the mixture. 
As a consequence, Eq.~(\ref{Eq: derivation of the prior distribution_step1}) can be written as 
\begin{equation}
\label{Eq: derivation of the prior distribution_step2}
b_{k,d,j} = 1+\frac{ \left( 1-\lambda_{k,d}(\bm{\eta}^{T-1}) \right) \cdot \sum_{i=1}^N  w_i }{K \cdot n_d}; \quad \quad j = 1,...,n_d.
\end{equation}
In general, both the relative strength of the data, $\lambda_{k,d}(\bm{\eta}^{T-1})$, and the sum of the weights, $\sum_{i=1}^N w_i$, increase with the sample size $N$, and we replace $\left( 1-\lambda_{k,d}(\bm{\eta}^{T-1}) \right) \cdot \sum_{i=1}^N  w_i$ by a constant $C$ in Eq.~(\ref{Eq: derivation of the prior distribution_step2}), which gives
\begin{equation}
\label{Eq: prior parameters_b}
b_{k,d,j} = 1 + \frac{C}{K \cdot n_d}; \quad \quad \quad j = 1,...,n_d
\end{equation}
for each $\bm{\theta}_{k,d}$. We will compare different choices of $C$ in the numerical examples.
As for the mixture weights $\bm{\alpha}$, we choose
\begin{equation}
\label{Eq: prior parameters_a}
a_{k} = 1 + \epsilon; \quad \quad \quad k = 1,...,K,
\end{equation}
where $\epsilon$ is typically set as a small value, e.g. $10^{-8}$. 

In fact, we penalize the weighted log-likelihood with the following log-prior term
\begin{equation}
\ln\left( p(\bm{\eta}|\mathcal{M}_k ) \right) 
= \ln \text{Dir}(\bm{\alpha}|\bm{a}) + \sum_{k=1}^K\sum_{d=1}^D \ln \text{Dir}(\bm{\theta}_{k,d} | \bm{b}_{k,d}).
\end{equation}
For symmetric Dirichlet distributions $\text{Dir}(\bm{\alpha}|\bm{a})$ and $\text{Dir}(\bm{\theta}_{k,d})$ defined in Eq.~(\ref{Eq: prior parameters_b}) and (\ref{Eq: prior parameters_a}), the probability mode is attained when $\alpha_k = 1/K, k=1,...,K$ and $\theta_{k,d,j}= 1/n_d, j = 1,...,n_d$. 
In other words, we favor a uniform vector for each $\bm{\theta}_{k,d}$, and a larger $C$ implies a stronger preference.
Note that by selecting a small $\epsilon$, the penalization of non-uniform $\bm{\alpha}$ vanishes, so the redundant mixture components can be assigned a small weight. 
\paragraph{Model selection or not} 
To discuss whether it is necessary to perform model selection, we consider two categorical mixtures $f_{m1}(\cdot|\bm{\eta}, \mathcal{M}_{K1})$ and $f_{m2}(\cdot|\bm{\eta}, \mathcal{M}_{K2})$. 
Let $K1>K2$ and we refer to $f_{m1}$, $f_{m2}$ as the larger mixture, and the smaller mixture, respectively. 
Through, for example, adding $K1-K2$ redundant mixture components, each of zero weight to $f_{m2}$, any distribution that can be represented by the smaller mixture $f_{m2}$ can also be represented by the larger one $f_{m1}$.
Therefore, the minimum KL divergence between the optimal IS distribution and the larger mixture will be less or equal to that of the smaller mixture, and if we can always find the optimal parameters $\bm{\eta}^*$ defined in Eq.~(\ref{Eq: CE optimization for intermediate target dist. in iCE}), the BiCE-CM with a larger mixture will perform better or at least equally well than using a smaller mixture.

If the sample size approaches infinity, the distribution implied by either the weighted MLE $\widehat{\bm{\eta}}$ or the weighted MAP $\widetilde{\bm{\eta}}$ converges to the distribution implied by the optimal parameters $\bm{\eta}^*$, and if we can always find the weighted MLE or weighted MAP through the generalized EM algorithm, there is no need to perform model selection, since the larger the $K$, the closer the chosen IS distribution is to the optimal IS distribution, and thus the better the performance of the CE method.
    
In practical settings, the sample size is limited, and the weighted MLE $\widehat{\bm{\eta}}$ can be far away from the optimal parameter $\bm{\eta}^*$. 
Although by introducing the prior information, the overfitting issue of the weighted MLE is mitigated, there is still no guarantee that the distribution implied by the weighted MAP $\widetilde{\bm{\eta}}$ is close to that of $\bm{\eta}^*$. 
Even if the weighted MAP of a mixture can be found, it does not necessarily lead to a closer distribution to the optimal IS distribution than using the weighted MAP of a smaller mixture, especially when an inappropriate prior distribution is chosen, and hence, we cannot simply employ a large $K$.
 
Another major issue is that in practice the generalized EM algorithm almost always gets stuck at a local maximum and fails to identify the weighted MAP.
Note that there are in total $K^n$ terms (usually uni-modal) in the likelihood function. 
Although some of these terms can be merged, a large sample size $n$ or number of mixture components $K$ generally indicates a more complicated and jagged posterior surface, whereby our generalized EM optimizer is more likely to get stuck at a point far from optimal. 
In such cases, a higher effort is required to find a good local maximum, e.g., by launching more pilot runs or designing a special prior that eliminates some of the modes.

In summary, it is challenging to make a general decision on whether or not to perform the model selection, and we select the $K$ with the highest posterior probability among a set of $K_{max}$ candidates. 
The posterior probability can be roughly approximated by twice the negative BIC in Eq.~(\ref{Eq: BIC}).
Although such an approximation suffers from major limitations, it remains one of the state-of-art techniques for selecting the number of components in a mixture model. 
For more details, we refer to Sec.~\ref{Subsec: Model selection and BIC}.


\paragraph{The algorithm}
The proposed generalized EM algorithm for inference of the categorical mixture is summarized in Algorithm~\ref{Alg: gEM}.
\begin{algorithm}[htbp]
\caption{The generalized EM algorithm}
\label{Alg: gEM}
\small
\SetKwProg{algo}{MainFunc}{:}{}
\algo{}{
\LinesNumbered
\KwIn{$\{ \bm{x}_i, W_i \triangleq W(\bm{x}_i) \}_{i=1}^N$, $C$, $\epsilon$, $K$, $\Omega_{\bm{X}} \triangleq \{ s_{d,1},..., s_{d,n_d}\}_{d=1}^D$ }

\% $\Omega_{\bm{X}}$ is the sample space of $\bm{X}$, $W(\cdot)$ is defined by Eq.~(\ref{Eq: weights in iCE})

$w_i \leftarrow N \cdot \frac{W_i}{\sum_{i=1}^N{W_i}}$ for each $i=1,...,N$ \% normalizing the weights \\

$n_p \leftarrow 20$ \% the number of the pilot runs\\
$l_p \leftarrow 20$ \% the maximum iteration of the pilot run\\
$l_o \leftarrow 500$ \% the maximum iteration of the official run\\
$\mathcal{LP}_{max} \leftarrow -\infty $ \% the maximum weighted log-posterior of the pilot runs\\

$it \leftarrow 1$ \% the counter for the pilot run\\
\While{$it \leqslant n_p$}{
\For{$i=1,...,N$}{ 
Generate $\{ \gamma^{(0,it)}_{i,k} \}_{k=1}^K$ uniformly over the standard $(K-1)$ simplex
} 
$\left( \sim, \mathcal{LP}, \sim \right) = \textbf{Subroutine} \left( \{ \bm{x}_i, w_i \}_{i=1}^N,  [\gamma^{(0,it)}_{i,k}]_{N \times K}, \Omega_{\bm{X}}, C, \epsilon, l_p \right)$ \\
\If{$\mathcal{LP} \geqslant \mathcal{LP}_{max}$}{
$ \gamma^{(0)}_{i,k} \leftarrow \gamma^{(0,it)}_{i,k}$ for each $i$ and $k$, 
$\mathcal{LP}_{max} \leftarrow \mathcal{LP}$
}
$it = it + 1$
} 
$\left( \mathcal{LL}, \mathcal{LP}, \widetilde{\bm{\mu}}_K \right) = \textbf{Subroutine} \left( \{ \bm{x}_i, w_i \}_{i=1}^N,  [\gamma^{(0)}_{i,k}]_{N \times K}, \Omega_{\bm{X}}, C, \epsilon, l_o \right)$ \\
Compute $BIC_K$ through Eq.~(\ref{Eq: BIC})\\
\KwOut{$\widetilde{\bm{\mu}}_K$, $BIC_K$}
}
\SetKwProg{subroutine}{Subroutine}{:}{}
\subroutine{}{
\setcounter{AlgoLine}{0}
\KwIn{$ \{ \bm{x}_i, w_i \}_{i=1}^N,  [\gamma_{i,k}]_{N \times K}, \Omega_{\bm{X}}, C, \epsilon, t_{max} $}
$tol \leftarrow \frac{1}{10 \cdot N}$, $r \leftarrow \infty$, $t \leftarrow 1$, $\mathcal{LP}^{(0)} \leftarrow 1$ \\
\While{$ r \geqslant tol $ and $t \leqslant t_{max}$}{
M step: plug $\gamma_{i,k}$, $w_i$ and $\Omega_{\bm{X}}$ into Eq.~(\ref{Eq: weighted MAP for alpha}) and (\ref{Eq: weighted MAP for theta}) and compute $\alpha_k$ and $\theta_{k,d,j}$ with $a_k$ and $b_{k,d,j}$ defined in Eq.~(\ref{Eq: prior parameters_a}) and (\ref{Eq: prior parameters_b}), respectively\\
E step: update $\gamma_{i,k}$ through Eq.~(\ref{Eq: optimal distribution of Z_i}). \\
compute the weighted log-likelihood $\mathcal{LL}^{(t)}$ and the weighted log-posterior $\mathcal{LP}^{(t)}$ via Eq.~(\ref{Eq: weighted log-posterior}) \\
$r \leftarrow \frac{ |\mathcal{LP}^{(t)}-\mathcal{LP}^{(t-1)}| }{\mathcal{LP}^{(t-1)}}$, $t \leftarrow t + 1$
}
let $\widetilde{\bm{\mu}}$ collect all $\alpha_k$ and $\theta_{k,d,j}$ \\ 
$\mathcal{LL} \leftarrow \mathcal{LL}^{(t-1)}$, $\mathcal{LP} \leftarrow \mathcal{LP}^{(t-1)}$\\
\KwOut{$\mathcal{LL}$, $\mathcal{LP}$, $\widetilde{\bm{\mu}}$}
}
\end{algorithm}

\subsection{Bayesian improved cross entropy method with the categorical mixture model}
\label{Subsec: BiCE with CatMix}
The BiCE method \cite{Chan&others2023} substitutes the weighted MLE of model parameters in the original iCE method with a Bayesian counterpart. 
In \cite{Chan&others2023}, the posterior predictive distribution is derived for updating the independent categorical distribution. 
However, for the categorical mixture, a closed-form expression of the posterior predictive distribution does not exist, and we use the weighted MAP estimator instead, which can be approximated through a generalized EM algorithm described in Subsec.~\ref{Subsec: MAP for CatMix in iCE}.
The proposed BiCE method with the categorical mixture model (BiCE-CM) is summarized in Algorithm~\ref{Alg: BiCE-CM}.
\begin{algorithm}[htbp]
\caption{Bayesian improved cross entropy method with the categorical mixture parametric family}
\label{Alg: BiCE-CM}
\LinesNumbered
\KwIn{$N$, $\delta_{tar}$, $\delta_{\epsilon}$, $C$, $\epsilon$, the maximum number of mixture components $K_{max}$, performance function $g(\bm{x})$, input distribution $p_{\bm{X}}(\bm{x})$, $\bm{x} \in \Omega_{\bm{X}}$ } 

$t \gets 1$, $t_{max} \gets 50$, $\sigma_0 \gets \infty$\\
$h(\bm{x}; \widetilde{\bm{\mu}}^{(t-1)}) \gets p_{\bm{X}}(\bm{x})$\\ 
\While {true}{ 
	Generate $N$ samples $\{ \bm{x}_k \}_{k=1}^{N}$ from $h(\bm{x}; \widetilde{\bm{\mu}}^{(t-1)})$ and calculate the corresponding performance $\{g(\bm{x}_k)\}_{k=1}^{N}$\\
	Compute the sample c.o.v. $\widehat{\delta}$ of $\left\{ \frac{\mathbb{I}\{ g(\bm{x}_k)\leq0 \}}{\Phi(-g(\bm{x}_k)/\sigma^{(t-1)})} \right\}_{k=1}^{N}$\\	 
	\If{ $t>t_{max} $ or $ \widehat{\delta} \leq \delta_{\epsilon}$}{Break
	} 
	Determine $\sigma^{(t)}$ through solving Eq.~(\ref{Eq: the updating rule of sigma}) using the alternative weight function $W^{(alt)}(\cdot)$ defined in Eq.~(\ref{Eq: weights in BiCE})\\
	Calculate $W(\bm{x}_i)$ for each $i=1,...,N$ through Eq.~(\ref{Eq: weights in iCE})\\
	\For{$K = 1,...,K_{max}$}{
	Compute $\widetilde{ \bm{\mu} }_K$ and $BIC_K$ through Algorithm~\ref{Alg: gEM} 
	}
	$\widetilde{K} = \argmin_{K} BIC_K$\\
	$\widetilde{\bm{\mu}}^{(t)} \leftarrow \widetilde{ \bm{\mu} }_{\widetilde{K}}$\\
	$t \gets t+1$
}
$T \gets t-1$\\
Use $h(\bm{x}; \widehat{\bm{v}^{(T)}})$ as the IS distribution and calculate the IS estimator $\widehat{p}_f$\\
\KwOut{$\widehat{p}_f$}
\end{algorithm}
\subsection{Component importance measures from the BiCE-CM algorithm}
\label{Subsec: CI from BiCE-CM}   
In the field of network reliability assessment, component importance (CI) measures are employed for ranking components based on their influence on the system failure probability. 
Commonly used CI measures for binary systems include among others Birnbaum's measure, critical importance factor, risk achievement worth, and Fussel-Vesely  measure \cite{Rausand&Hoyland2003}. 
These measures can be extended to multi-state or continuous systems \cite{Ramirez-marquez&others2005}, e.g., after introducing a performance function $g_i(\cdot)$ at the component level \cite{Zio&Podofillini2003}, i.e., the $i$-th component fails when $g_i(x_i)\leqslant 0$. 

The samples from the BiCE-CM method can be used for calculating these CI measures. 
Taking Birnbaum's measure (BM) as an example, it is defined as the partial derivative of the system failure probability $p_f\triangleq \Pr(g(\bm{X})\leqslant 0)$ with respect to the component failure probability $p_{fi}\triangleq \Pr( g_i(X_i) \leqslant 0 )$:
\begin{align}
\label{Eq: BM}
BM_i 
& \triangleq \frac{\partial p_f}{\partial p_{fi}} = \Pr( g(\bm{X})\leqslant 0| g_i(X_i)\leqslant 0) - \Pr( g(\bm{X})\leqslant 0| g_i(X_i)> 0) \notag \\
& = \frac{\Pr( g(\bm{X})\leqslant 0, g_i(X_i)\leqslant 0)}{\Pr(g_i(X_i)\leqslant 0)} - \frac{\Pr( g(\bm{X})\leqslant 0, g_i(X_i) > 0)}{\Pr(g_i(X_i) > 0)} \notag \\
& = \frac{ \mathbb{E}_{p_{\bm{X}}} \left[ \mathbb{I}\{ g(\bm{X}) \leqslant 0\} \mathbb{I}\{ g_i(X_i) \leqslant 0\} \right] }{p_{fi}} - \frac{\mathbb{E}_{p_{\bm{X}}} \left[ \mathbb{I}\{ g(\bm{X}) \leqslant 0\} \mathbb{I}\{ g_i(X_i) > 0\} \right]}{1-p_{fi}}. 
\end{align}   
The expectation in Eq.~(\ref{Eq: BM}) can be estimated through IS using the samples from the final level of the BiCE-CM method, and $p_{fi}$ can be estimated by crude MCS with $g_i(X_i)$, which is usually cheap to evaluate. 
According to the definition, the larger the $BM_i$, the more sensitive the failure probability $p_f$ is to the $i$-th component, and hence the higher priority the component will have when allocating the system redundancy.     
   
\section{Numerical examples}
\label{sec: Examples}
\subsection{Illustration: a toy connectivity problem}
\begin{figure}[htbp]
    \centering
	\includegraphics[scale=0.8]{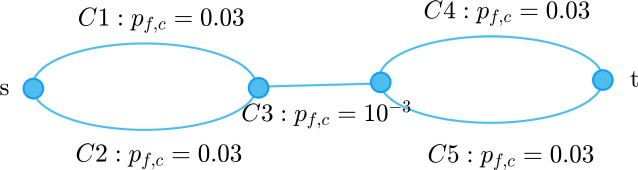}
    \caption{Topology of a five-component network in Example 5.1.}
	\label{Fig: TopologyOfAsmallNetwork}
\end{figure}

We consider a small network consisting of five components. 
Its configuration is shown in Fig.~\ref{Fig: TopologyOfAsmallNetwork}. 
Each component can either fail or not fail and hence is modeled by a Bernoulli distributed random variable.
The topologically most important component, component 3, is assigned a failure probability of $10^{-3}$, while for all other components, the failure probability is set to $3\cdot10^{-2}$. 
The connectivity between points A and B is of interest, and we have three major modes in the failure domain: $(0,0,1,1,1)$, $(1,1,0,1,1)$, and $(1,1,1,0,0)$, corresponding to three minimal cut sets: $(1,2), (3)$, and $(4,5)$, respectively. 
The probability of each mode equals $8.46\cdot 10^{-4}, 8.85\cdot10^{-4}, 8.46\cdot 10^{-4}$, respectively, and the total failure probability equals $2.80 \cdot 10^{-3}$. 
\subsubsection{The zero count problem for the iCE}
\begin{figure}[htbp]
    \centering
	\includegraphics[scale=0.7]{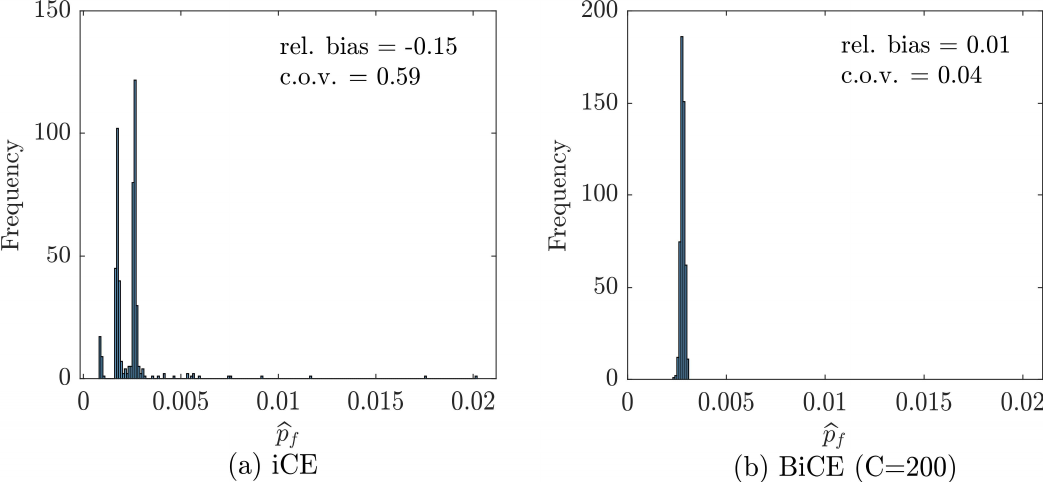}
    \caption{Histogram of the failure probability estimates via the iCE or the BiCE-CM method. $(a)$ results of the iCE. $(b)$ results of the BiCE-CM.}
	\label{Fig: Histogram_CoonOfAsmallNetwork_tarCV1_N1000_K3_iCEtoBiCE}
\end{figure}

To illustrate the overfitting issue of the standard iCE method when solving this example, we run it 500 times with the setting $K=3, \delta_{tar}=\delta_{\epsilon}=1, N=1000$ and plot the histogram of the 500 failure probability estimates in Fig.~\ref{Fig: Histogram_CoonOfAsmallNetwork_tarCV1_N1000_K3_iCEtoBiCE}. 

The figure illustrates a highly skewed but also multi-modal distribution of the iCE estimator. The three peaks reflect the number of cases where zero, one, or two modes are missing in the final IS distribution.
A 'missing' mode here implies that the mode is assigned a small (even zero) probability by the IS distribution. 
Any sample coincides with such a mode will be attached with a large weight, leading to an outlier that significantly overestimates the failure probability.
By contrast, if no sample is generated from this mode, there will be a significantly negative bias. 
Note that the number of samples from the nominal distribution whose third component is safe follows a binomial distribution and therefore its properties can be calculated theoretically. 
For instance, the probability that the third component is safe for all samples generated at the first level is equal to $(1-10^{-3})^{1000} \approx 0.368$.  
In such case, the iCE method will definitely miss the mode $(3)$ in all subsequent intermediate levels (see Eq.~(\ref{Eq: updating rule of the theta})), and the corresponding failure probability estimates will underestimate the true value, which is demonstrated in Fig.~\ref{Fig: Histogram_CoonOfAsmallNetwork_tarCV1_N1000_K3_iCEtoBiCE}. 

In Fig.~\ref{Fig: Histogram_CoonOfAsmallNetwork_tarCV1_N1000_K3_iCEtoBiCE}(b), we show the results for the BiCE-CM method. 
A balanced Dirichlet prior in Eq.~(\ref{Eq: prior parameters_b}) is chosen for mixture parameters, with $C=200$ and $\epsilon = 10^{-8}$. 
The remaining settings are the same as those of the iCE method. 
We can see that by introducing an appropriate prior, all three modes are found in most of the 500 estimates. 
A negligible relative bias (0.45\%) and a small coefficient of variation (0.1) are achieved with an average of 4050 evaluations of $g(\cdot)$.

\begin{figure}[htbp]
    \centering
	\includegraphics[scale=0.7]{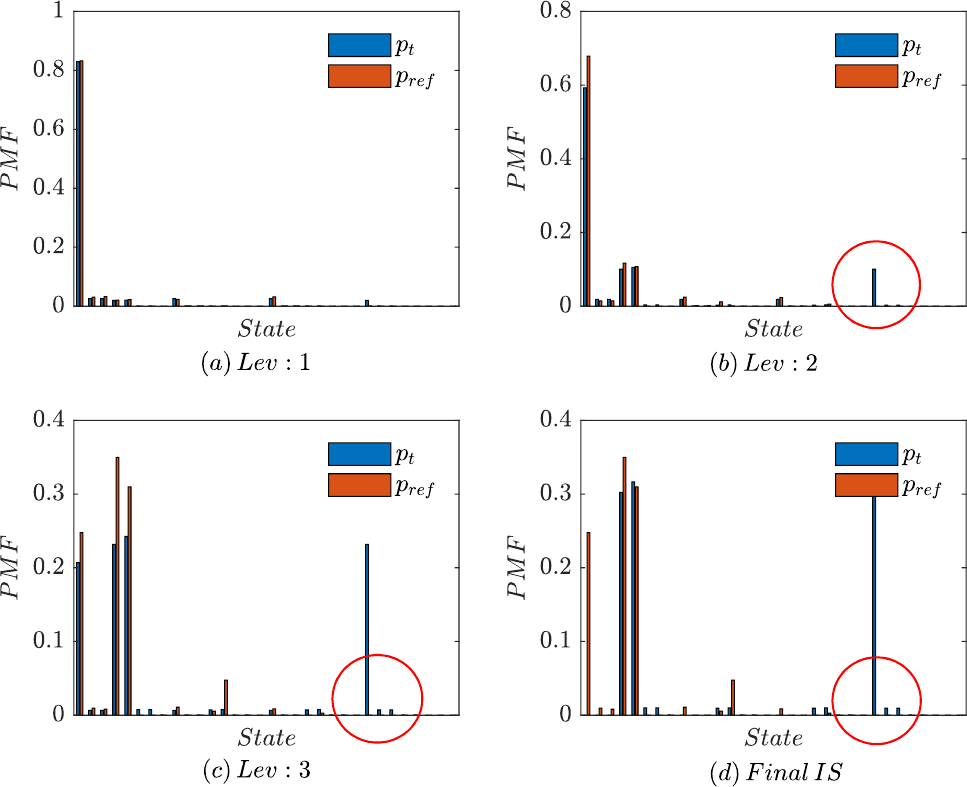}
    \caption{The PMF of the target distribution and of the reference distribution at each iteration of the iCE method.}
	\label{Fig: IntermidateResults_ConnOfAsmallNetwork}
\end{figure}
To investigate the reason for the significant difference between the performance of the two algorithms, we keep track of the reference distributions of all intermediate levels of the iCE method. 
The results are shown in Fig.~\ref{Fig: IntermidateResults_ConnOfAsmallNetwork}. 
Fig.~\ref{Fig: IntermidateResults_ConnOfAsmallNetwork}(a) demonstrates whether the distribution chosen at each level of the iCE method, i.e., the reference distribution, resembles the target distribution well. 
Apparently, the iCE method misses one of the three modes in the optimal IS distribution starting from the second level and produces a biased estimate. 
\subsubsection{Model selection or not: an empirical perspective}
Next, we use the BIC for choosing adaptively the number of mixture components $K$ at each level of the BiCE-CM. 
The maximum number of mixture components $K_{max}$ is equal to 10.
For comparison, we also perform the BiCE-CM method with a fixed number of $K$ ranging from 1 to 100. 
Overall, 8 scenarios are considered as listed in Table~\ref{Table: case description}. 
In all cases, a Dirichlet distribution is employed as a priori with $C = 200$ and $\epsilon = 10^{-8}$, and  $\delta_{tar}=\delta_{\epsilon}$ is set to 1. 
\begin{table}[htbp]
    \centering
    \footnotesize
    \caption{Case description for example 5.1.2.} 
    \label{Table: case description}
    \begin{tabular}{lllllllll}
	\hline
    Case No. & 1 & 2 & 3 & 4 & 5 & 6 & 7 & 8\\  
    \hline       
    number of mixture components, $K$ & 1 & 2 & 3 & 5 & 10 & 20 & 100 & BIC\\       
    \hline  
    \end{tabular}
\end{table}
\begin{figure}[htbp]
    \centering
	\includegraphics[scale=0.7]{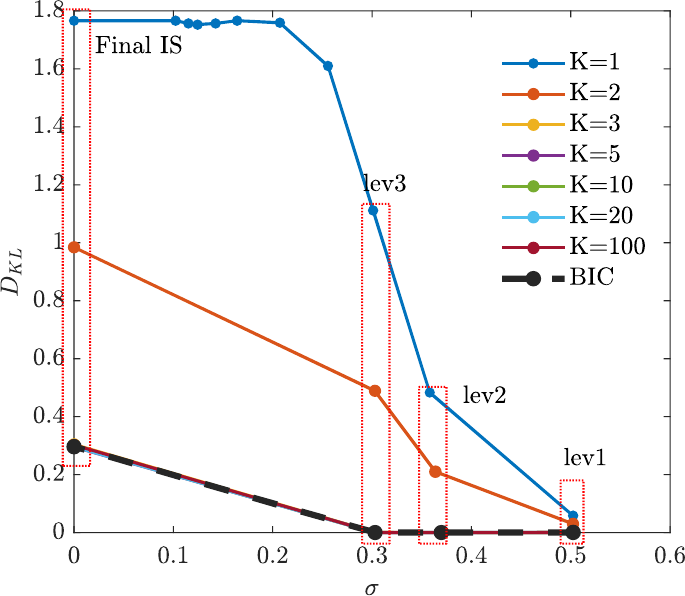}
    \caption{KL divergence between the intermediate target distribution and the reference distribution at each level of the BiCE-CM method (a large sample setting).}
	\label{Fig: ConnOfAsmallNetwork_KLdivergenceAtDifferentIntermediateLevel}
\end{figure}

We first consider a large sample setting, where $N=10^5$, and check the estimated KL divergence between the intractable target distribution and its mixture approximation, the reference distribution, at each level of the BiCE-CM method. 
The results are illustrated in Fig.~\ref{Fig: ConnOfAsmallNetwork_KLdivergenceAtDifferentIntermediateLevel}. 
We can see from the figure that the estimated KL divergence at each intermediate level decreases as $K$ increases, and reaches a constant minimum value at $K=3$. 
This result is expected since the optimal IS distribution has three major modes and can be approximated sufficiently well by a three-component categorical mixture. 
Hence, additional flexibility from adding mixture components is not required. 
However, for $K<3$, the model capacity is inadequate, and increasing $K$ will lead to an IS distribution significantly closer to the optimal one thus clearly improving the performance of the BiCE-CM. 
Fig.~\ref{Fig: ConnOfAsmallNetwork_KLdivergenceAtDifferentIntermediateLevel} also demonstrates that selecting the $K$ adaptively via BIC will not improve the results of a fixed $K$ that is larger than 3, so the model selection is not needed in large sample settings for this example. 

\begin{figure}[htbp]
    \centering
	\includegraphics[scale=0.65]{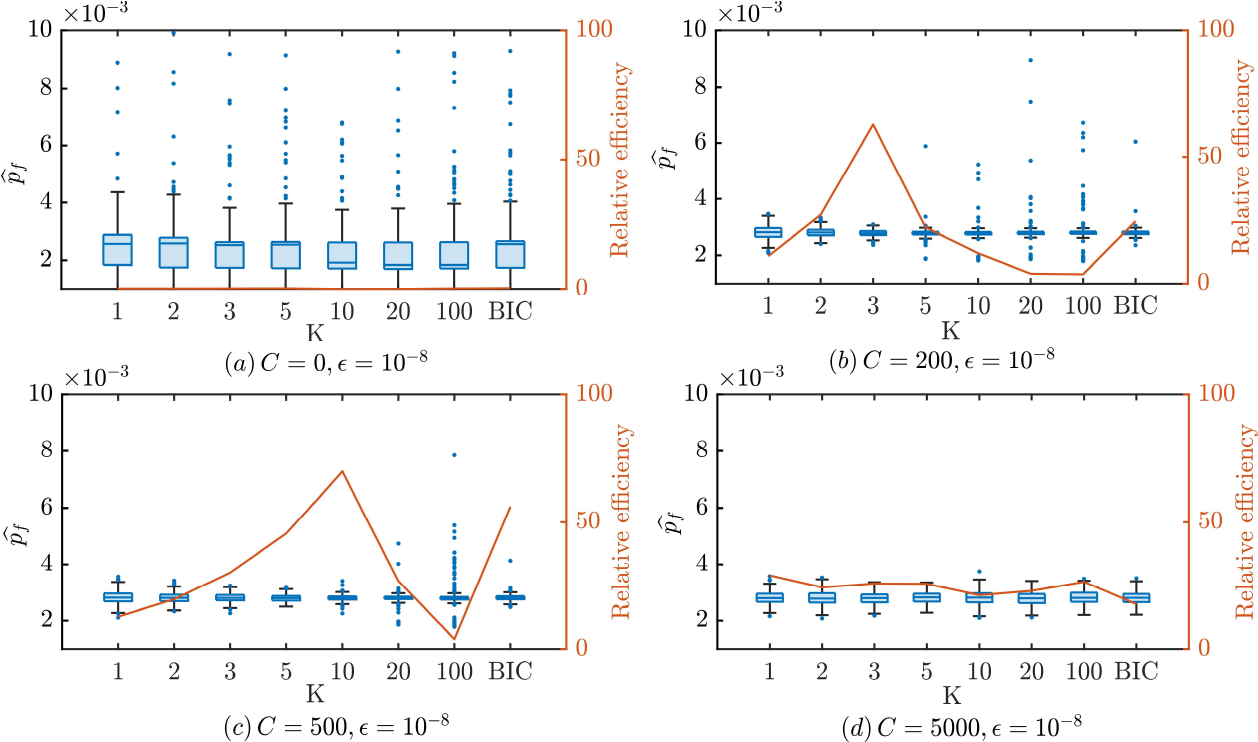}
    \caption{Boxplot of the estimates obtained from the BiCE-CM method. (a) $C=0, \epsilon =10^{-8}$, (b) $C=200, \epsilon =10^{-8}$, (c) $C=500, \epsilon =10^{-8}$, (d) $C=5000, \epsilon =10^{-8}$.}
	\label{Fig: ImpactOfDifferentPriorDist}
\end{figure}

Next, we consider small sample settings, in which the weighted MLE tends to overfit the data. 
Although introducing a prior distribution mitigates the overfitting issue for an appropriate choice of the prior parameters, such a choice is not always straightforward. 
That is, a poor parameter choice of the prior for a model with higher $K$ could potentially result in a worse estimator. Such situations can be avoided by performing model selection. 
This is demonstrated by the numerical experiment, where for each scenario we run 500 times the BiCE-CM algorithm with $1,000$ samples and we set $C=200, \epsilon=10^{-8}$. 
The results are summarized through a box plot in Fig.~\ref{Fig: ImpactOfDifferentPriorDist}(b).

To measure the quality of the failure probability estimator $\widehat{p}_f$, we borrow the definition of the 'efficiency' in statistics \cite{L'Ecuyer1994}, which is defined as follows   
\begin{equation}
\label{Eq: efficiency of an estimator}
\text{Eff}(\widehat{p}_f) \triangleq \frac{1}{\text{MSE}(\widehat{p}_f) \times \text{Cost}(\widehat{p}_f)},
\end{equation}
where $\text{MSE}(\widehat{p}_f)$ represents the mean square error of the estimator $\widehat{p}_f$ and $\text{Cost}(\widehat{p}_f)$ is the average computational cost of getting $\widehat{p}_f$, which is measured by the average number of evaluations of $g(\cdot)$ throughout all numerical examples in this paper. 
Note that the efficiency of the MCS equals $\frac{1}{p_f \cdot (1-p_f)} $, which is independent of the sample size. Hence, the efficiency improvement over MCS can be measured through the following relative efficiency
\begin{equation}
\label{Eq: relative efficiency of an estimator}
\text{relEff}(\widehat{p}_f) \triangleq \frac{p_f \cdot (1-p_f)}{\text{MSE}(\widehat{p}_f) \times \text{Cost}(\widehat{p}_f)}.
\end{equation}
The relative efficiency of different choices of $K$ is illustrated in Fig.~\ref{Fig: ImpactOfDifferentPriorDist}(b). 
The optimal choice, as expected, is $K=3$. 
If guessing an appropriate $K$ is not possible, adaptively selecting $K$ via the BIC can be a good alternative. 
Note that this comes at a price of a significant overhead, since at each iteration, the generalized EM algorithm is performed $K_{max}=10$ times, while for a fixed $K$, we only perform one single run of the algorithm. 
Nevertheless, for a computationally demanding performance function $g(\cdot)$, the computational cost is dominated by the evaluation of $g(\cdot)$ and the overhead resulting from the adaptive selection of $K$ via the BIC should not be critical.
\subsubsection{Impact of the prior distribution}
In this subsection, we study the influence of the prior distribution on the performance of the BiCE-CM method. 
We consider 4 different values of $C$, namely $0,200,500$ and $5,000$. 
$\epsilon$ is fixed at $10^{-8}$ for all 4 cases. 
The results are summarized in Fig.~\ref{Fig: ImpactOfDifferentPriorDist}. 
When $C=0$, the BiCE-CM method degenerates to the standard iCE method that employs the weighted MLE to update the mixture model. 
Due to overfitting, the relative efficiency is poor. 
When $C=5000$, the weighted log-likelihood function is over-penalized, and the prior estimate dominates the data-related estimate in Eq.~(\ref{Eq: rephrasing of the updating formula of theta}). 
Owing to the symmetric Dirichlet prior, the resulting IS distribution is close to an independent uniform distribution, and the BiCE-CM with different $K$ performs similarly. 
For this 5-component toy example, an independent uniform distribution works well, however, as will be shown later, this is not generally the case. 
When $C$ is appropriately large, the performance of the BiCE-CM method is shown in Fig.~\ref{Fig: ImpactOfDifferentPriorDist}(b-c), and has been discussed in Subsec.~5.1.2.    

\subsection{Comparison: a benchmark study}
In this subsection, we consider the multi-state two-terminal reliability problems \cite{Jane&Laih2008}, in which we compute the probability that a specified amount of 'flow' can (or cannot) be delivered from the source to the sink. 
This problem has been extensively studied in operations research \cite{Ramirez-Marquez&Coit2005, Jane&Laih2008, Yeh2015, Botev&others2018, Cancela&others2019}, from which we borrow two benchmark problems, namely the Fishman network and the Dodecahedron network, to test the performance of the BiCE-CM method. 
The results are further compared with the creation-process-based splitting (CP-splitting) \cite{Cancela&others2019}, which is a state-of-art technique for solving multi-state two-terminal reliability problems, especially when the failure probability $p_f$ is small.
  
\begin{figure}[htbp]
    \centering
	\begin{subfigure}[b]{0.45\linewidth}
		\includegraphics[width=\linewidth]{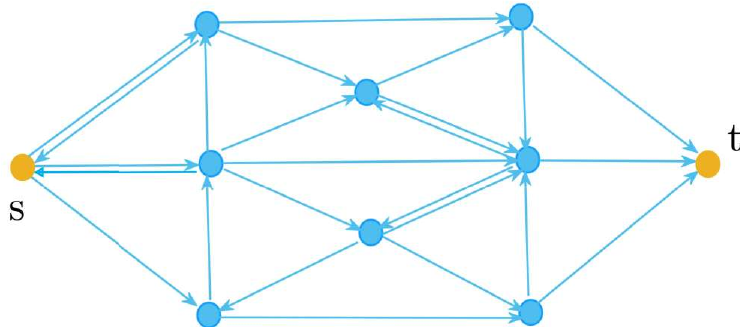}
		\caption{Fishman network.}
	\end{subfigure}
	\begin{subfigure}[b]{0.45\linewidth}
		\includegraphics[width=\linewidth]{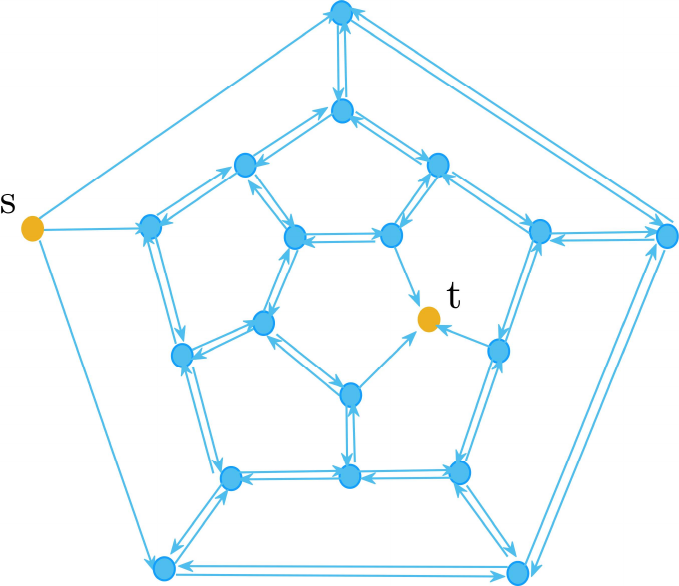}
		\caption{Dodecahedron network.}
	\end{subfigure}
	\caption{Topology of the two benchmarks in Example 5.2.}
	\label{Fig: Topo. of the two benchmarks}
\end{figure}

The network topology of the two benchmarks is illustrated in Fig \ref{Fig: Topo. of the two benchmarks}, and we employ the same problem settings as in \cite{Cancela&others2019}. 
We consider only the edge capacities, each following an independently and identically distributed categorical distribution. 
Following this distribution the probability of each edge capacity being $0, 100, 200$ equals $p_0, \frac{1-p_0}{2}, \frac{1-p_0}{2}$ respectively. 
We are interested in the probability that the maximum flow from the source node $s$ to the sink node $t$ is less or equal to the threshold $thr$, i.e., $\Pr(\text{mf}(s,t) \leqslant thr)$. 
We estimate this probability for each combination of $p_0 \in \{ 10^{-3}, 10^{-4} \}$ and $thr \in \{0, 100\}$, and for each of the two benchmarks. 
The reference failure probability $p_{ref}$ in each scenario is calculated using the CP-splitting method with $10^6$ trajectories. 
The results are summarized in Table~\ref{Table: BiCE for Fishman network} and \ref{Table: BiCE for Dodecahedron network}.
 
For the BiCE-CM method, we set $N=2000, \delta_{tar}=\delta_{\epsilon}=1.5, C=200, \epsilon = 10^{-8}$, and compute the mean value, c.o.v., the average number of evaluations of $g(\cdot)$, and the relative efficiency through 500 independent repetitions of the algorithm. 
For the CP-splitting method, we report the results from Tables~3 and 4 in \cite{Cancela&others2019}. 
Therein, the c.o.v. is computed for the mean value of 1000 repetitions. 
To obtain the c.o.v. of a single repetition, which guarantees a fair comparison between the two methods, the c.o.v. reported in \citep{Cancela&others2019} is multiplied by $\sqrt{1,000}$. 
In addition, the number of $g(\cdot)$ evaluations in CP-splitting is computed by multiplying the number of levels by the number of trajectories, without considering the pilot run. 

The performance of the BiCE-CM method for the two benchmarks is demonstrated in Table~\ref{Table: BiCE for Fishman network} and \ref{Table: BiCE for Dodecahedron network}, in which the results of the CP-splitting method are enclosed in the parentheses for comparison. 
\begin{table}[h!]
    \centering
    \footnotesize
    \caption{Performance of the BiCE method for the Fishman network in example 5.2.} 
    \label{Table: BiCE for Fishman network}
    \begin{tabular}{llllllll}
	\hline		
	 & $p_{ref}$ & mean & c.o.v. & cost & relEff\\	
	\hline    
    $p_0:10^{-3},thr:100$ & $3.00 \cdot 10^{-6}$ & $3.03(3.00^*) \cdot 10^{-6}$ & $0.05(0.17)$ & $1.03(0.90) \cdot 10^4$ & $ 1.2(0.13) \cdot 10^{4}$\\ 
     $p_0:10^{-4},thr:100$ & $ 3.00 \cdot 10^{-8}$ & $3.01(3.00) \cdot 10^{-8}$ & $0.04(0.21)$ & $1.40(1.30) \cdot 10^4$ & $1.5(0.058) \cdot 10^{6}$\\       
     $p_0:10^{-3},thr:0$ & $2.03 \cdot 10^{-9}$ & $2.01(2.02) \cdot 10^{-9} $ & $0.04(0.24)$ & $1.40(1.40) \cdot 10^4$ & $2.1(0.062) \cdot 10^{7}$\\              
    $p_0:10^{-4},thr:0$ & $2.00 \cdot 10^{-12}$ &  $2.00(2.00) \cdot 10^{-12} $ & $0.04(0.28)$ & $1.80(1.80) \cdot 10^4$ & $1.7(0.035) \cdot 10^{10}$\\  
    \hline  
    \end{tabular}
    \vspace{0.1cm}
    \begin{minipage}{15cm}
    \tiny    
    \vspace{0.1cm}
    $^*$ The number in the parentheses shows the result of the CP-splitting method.
    \end{minipage}
\end{table}
\begin{table}[h!]
    \centering
    \footnotesize
    \caption{Performance of the BiCE method for the Dodecahedron network in example 5.2.} 
    \label{Table: BiCE for Dodecahedron network}
    \begin{tabular}{llllll}
	\hline	
	 & $p_{ref}$ & mean & c.o.v. & cost & relEff\\	
	\hline    
    $p_0:10^{-3},thr:100$ & $3.05 \cdot 10^{-6} $ & $3.04(3.03^*) \cdot 10^{-6}$ & $0.06(0.20)$ & $1.11(0.90) \cdot 10^4$ & $8.2(0.92) \cdot 10^{3}$\\ 
     $p_0:10^{-4},thr:100$ & $3.08\cdot 10^{-8} $ & $3.00(2.99) \cdot 10^{-8}$ & $0.06(0.23)$ & $ 1.40(1.30) \cdot 10^4$ & $7.6(0.49) \cdot 10^{5}$\\  
     $p_0:10^{-3},thr:0$ & $2.06 \cdot 10^{-9}$ & $2.01(2.03) \cdot 10^{-9} $ & $0.05(0.26)$ & $1.41(1.30) \cdot 10^4$ & $1.2(0.057) \cdot 10^{7}$\\          
    $p_0:10^{-4},thr:0$ & $2.02 \cdot 10^{-12}$ & $1.99(1.97) \cdot 10^{-12}$ & $0.06(0.27)$ & $1.80(2.10) \cdot 10^4$ & $7.4(0.34) \cdot 10^{9}$\\  
    \hline  
    \end{tabular}
    \vspace{0.1cm}
    \begin{minipage}{15cm}
    \tiny    
    \vspace{0.1cm}
    $^*$ The number in the parentheses shows the result of the CP-splitting method.
    \end{minipage}
\end{table}

From these two tables, we observe a clear variance reduction in the BiCE-CM estimator without increasing the computational cost compared to the CP-splitting method. 
The standard iCE performs poorly for these two benchmarks due to the choice of a small $p_0$. 

\begin{figure}[htbp]
    \centering
	\includegraphics[scale=0.7]{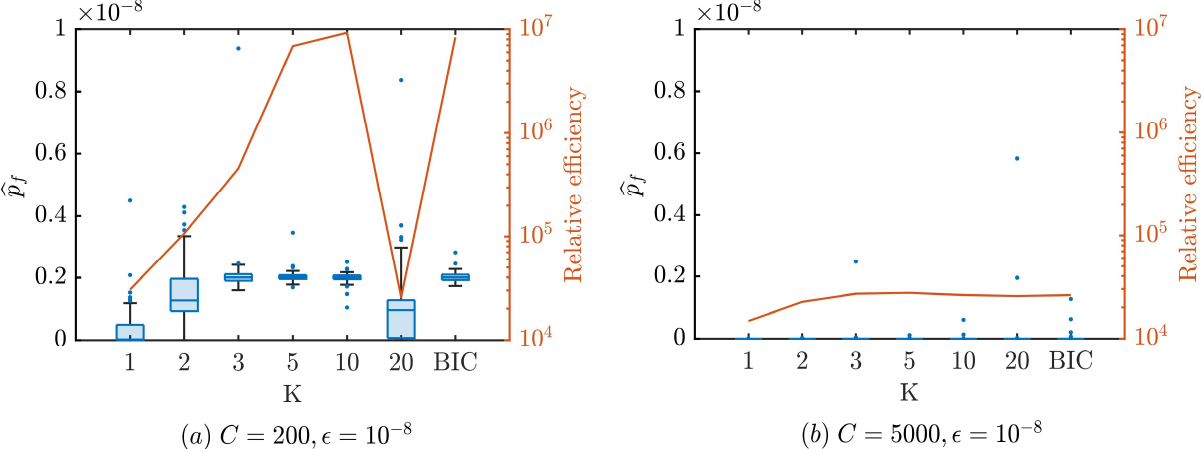}
    \caption{Boxplot of the BiCE-CM estimates for the Dodecahedron network with $thr=0$ and $p_0=10^{-3}$. (a) $C=200, \epsilon =10^{-8}$, (b) $C=5000, \epsilon =10^{-8}$.}
	\label{Fig: ImpactOfDifferentPriorDist_5.2}
\end{figure}

Fig.~\ref{Fig: ImpactOfDifferentPriorDist_5.2} illustrates the impact of different prior parameters $C$ and of different $K$ on the performance of the BiCE-CM method. 
We herein consider the Dodecahedron network with $thr=0$ and $p_0=10^{-3}$. 
When $C=5000$, the prior estimate dominates the data-related estimate in Eq.~(\ref{Eq: weighted MAP for theta}) and results in a near uniform IS distribution. 
In such cases, the performance of the BiCE-CM is poor. 
On the contrary, when $C=200$, which is a minor proportion of the $N$, the BiCE-CM works well for $K$ equal to 5 or 10 or when employing BIC. 

\subsection{Application: the IEEE 30 benchmark model with common cause failure}
\begin{figure}[htbp]
    \centering
	\includegraphics[scale=0.7]{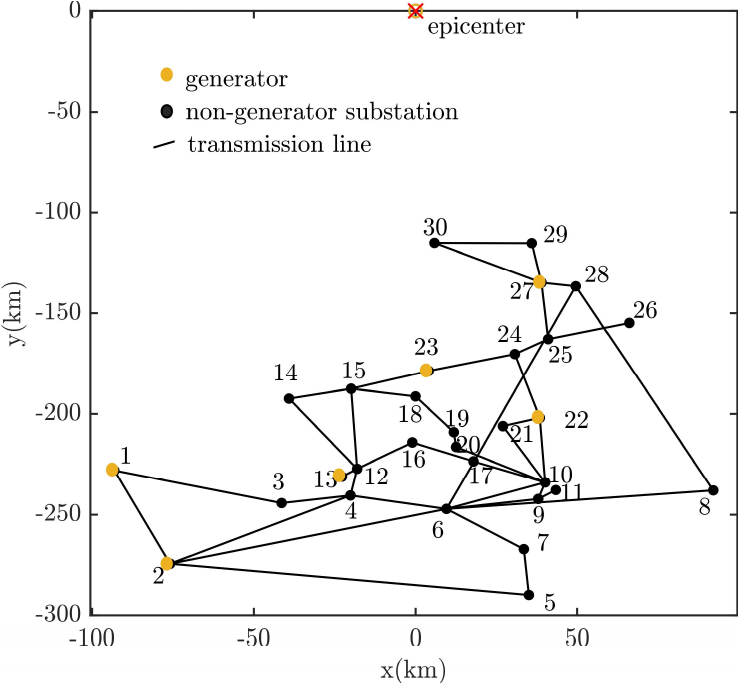}
    \caption{Network topology of the IEEE30 benchmark.}
	\label{Fig: Network topology of the IEEE30 benchmark}
\end{figure}

In this subsection, we consider the IEEE 30 power transmission network \cite{Zimmerman&others2010} illustrated in Fig.~\ref{Fig: Network topology of the IEEE30 benchmark}. 
The network consists of 6 power generators, 24 substations, and 41 transmission lines, which we assume to be subjected to earthquakes. 

\begin{figure}[htbp]
    \centering
	\includegraphics[scale=0.3]{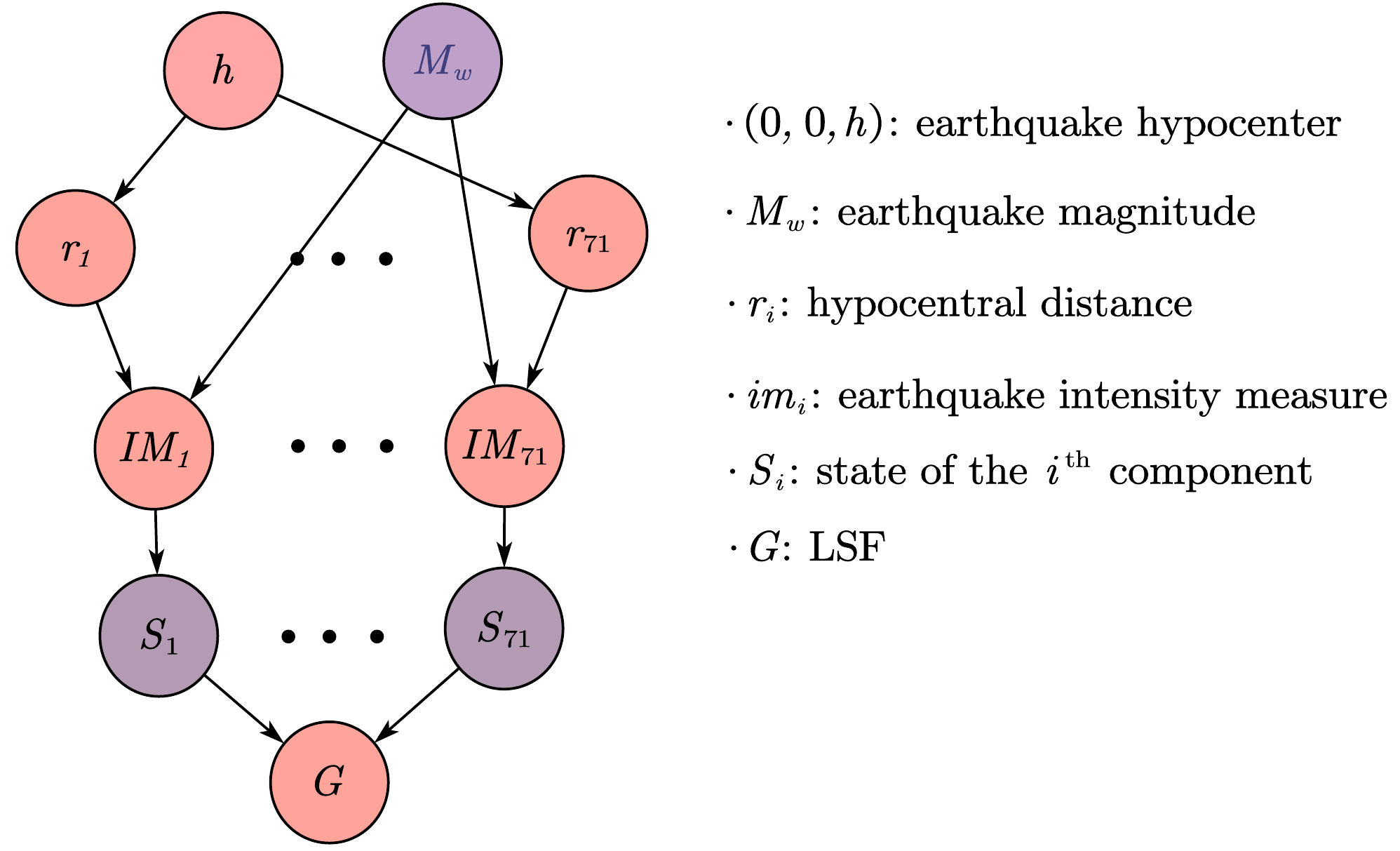}
    \caption{Dependence structure for the IEEE30 benchmark subjected to earthquakes. The purple nodes represent the random variables.}
	\label{Fig: BNforIEEE30}
\end{figure}

The hypocenter of the earthquake is assumed to be fixed and the earthquake magnitude is described by a truncated exponential distribution $p_M \propto  \exp(-0.85m), \quad 5 \leqslant m \leqslant 8$. 
The failures of the network components are dependent as they occur due to the earthquake, but it is often assumed that they are conditional independent given the earthquake \cite{Rosero-Velasquez&Straub2022}. 
Such conditional independence is depicted in Fig.~\ref{Fig: BNforIEEE30} \cite{Zwirglmaier&others2023}, where $r_i$ represents the hypocentral distance of the $i$-th component, and $im_i$ is the intensity measure of $i$. 
In the present example, $im_i$ is a deterministic function of $r_i$ described by the ground motion predictive equation (GMPE) given in \cite{Esteva&Villaverde1973}. 
$S_i$ denotes the state of the component $i$, whose distribution is indicated by the fragility curves in \cite{Cavalieri&others2014}. 
For each of the 6 generators, we consider 5 damage states, namely negligible, minor, moderate, extensive, and complete damage, which correspond to 0\%, 20\%, 60\%, 80\%, and 100\% reduction of power production, respectively.
The remaining 24 non-generator buses and all 41 transmission branches have 2 damage states, either safe or complete failure. 
The distribution of different network components is summarized in Table~\ref{Table: the dist. of diff. comp. for the IEEE30}.
\begin{table}[htbp]
    \centering
    \footnotesize
    \caption{The distribution of different components for the IEEE30 benchmark.} 
    \label{Table: the dist. of diff. comp. for the IEEE30}
    \begin{tabular}{llll}
	\hline
    & generators & non-generator buses &transmission lines\\  
    \hline
    \# components & 6 & 24 & 41\\
    distribution & categorical & Bernoulli & Bernoulli\\
	reference & Table~6.6 in [59] & Table~6.9 in [59] & $p_f = 5 \cdot 10^{-2}$\\ 
    \hline  
    \end{tabular}
\end{table}

\begin{figure}[htbp]
    \centering
	\includegraphics[scale=0.7]{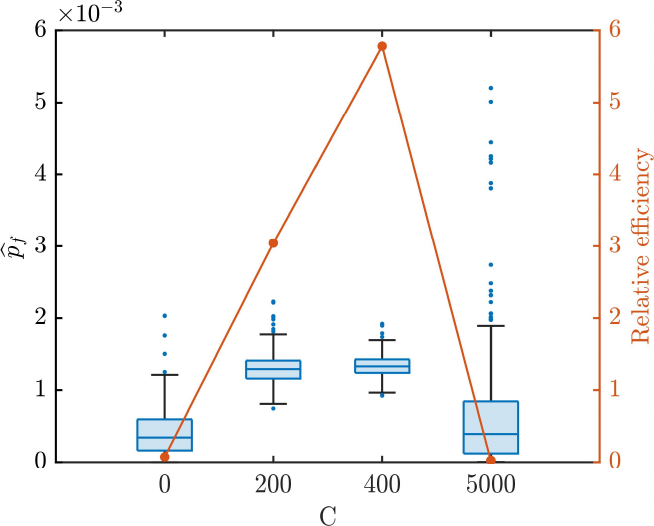}
    \caption{Boxplot of the BiCE-CM estimates for the IEEE30 benchmark model.}
	\label{Fig: ImpactOfDifferentPriorDist_5.3}
\end{figure}

We measure the network performance by the load shedding based on a direct current optimal power flow (DC-OPF) analysis using MATPOWER v7.1 \cite{Zimmerman&others2010}. 
The system failure is defined as over 50\% of the total power demand being shed after the earthquake, which gives the following performance function:
\begin{equation}
g(\bm{x}) \triangleq 50\% - \frac{LS(\bm{x})}{D_{tot}},
\end{equation}
where $LS(\bm{x})$ represents the load shedding with the network configuration, or state, $\bm{x}$, and $D_{tot}$ is the total power demand. 
The failure probability approximated by one single crude MCS with $10^6$ samples is equal to $0.0013$, which is then employed as the reference for validating the proposed BiCE-CM algorithm. 
For the BiCE-CM, 200 independent runs with $N=2,000, \delta_{tar}=\delta_{\epsilon}=1.5$ are launched, based on which, we calculate the mean, c.o.v. and the relative efficiency of the BiCE-CM estimator. 
The number of mixture components $K$ is adaptively chosen via the BIC, and we investigate 4 different prior distributions with $C \in \{0, 200, 400, 5000\}$ and $\epsilon = 10^{-8}$. 
The results are depicted in Fig.~\ref{Fig: ImpactOfDifferentPriorDist_5.3}, where it is shown that the BiCE-CM with $C=400$ performs the best among the four investigated cases. 
In particular, it significantly outperforms the $C=0$ case, which represents the standard iCE method.  
The relative efficiency of the BiCE-CM with $C=400$ is about 6, meaning the efficiency is around 6 times higher than that of the crude MCS. 
The average CPU time of the BiCE-CM is as $371.23$ seconds on a 3.50GHz Intel Xeon E3-1270v3 computer. 
As a comparison, crude MCS needs $46,161$ samples to achieve the same coefficient of variation as the BiCE-CM, which takes $1741.68$ seconds on the same computer. 
Hence, the overhead of BiCE-CM does not strongly affect the overall computation time. 

\begin{figure}[htbp]
    \centering
	\includegraphics[scale=0.7]{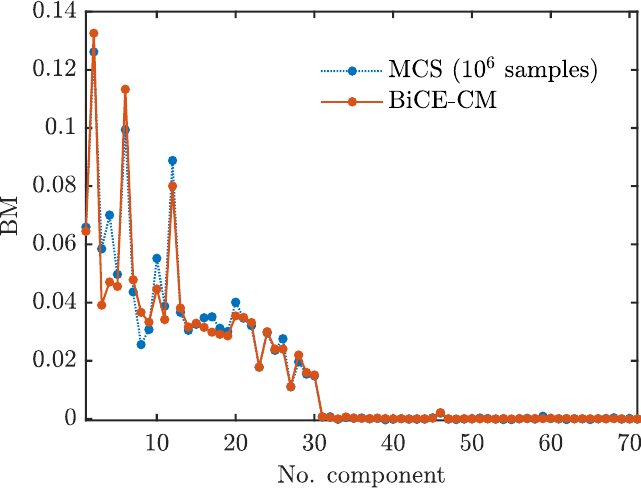}
    \caption{Birnbaum's measure for different components of the IEEE30 benchmark.}
	\label{Fig: Birnbaum's measure for different components of the IEEE30 benchmark.}
\end{figure}

The BM averaged over 200 repetitions of the BiCE-CM algorithm is depicted in Fig.~\ref{Fig: Birnbaum's measure for different components of the IEEE30 benchmark.} for different components of the IEEE30 benchmark model. 
For multi-state generators, the failure is defined as the power production being reduced by 80\% or more. 
We can see from the figure that except for components 3,4 and 8, the BM evaluated with the BiCE-CM method is consistent with that evaluated by crude MCS. 
\section{Conclusions}
In network reliability assessments, the network components are often strongly dependent given system failure. 
Such dependence cannot be captured by the independent categorical distribution employed in the original Bayesian improved cross entropy (BiCE) paper. 
To capture this dependence and improve the performance of the estimate, we employ instead the categorical mixture as the parametric family of the BiCE. 
The parameters of the mixture model are updated through the weighted maximum a posteriori (MAP) estimate. 
In this way, the overfitting issue encountered in the standard improved cross entropy (iCE) method, which employs the weighted maximum likelihood estimate (MLE), is mitigated. 
The proposed algorithm is termed the BiCE-CM method.

We approximate the weighted MAP through the expectation maximization(EM) algorithm with a minor modification to account for the weights and the prior. 
The algorithm results in a monotonically increasing weighted posterior and converges to a local maximum, a saddle point, or a boundary point depending on the starting point of the generalized EM algorithm. 
Moreover, the Bayesian information criterion (BIC) can be computed as a by-product of the generalized EM algorithm and is employed as model selection technique for choosing the optimal number of components in the mixture when the sample size is moderate. 
The model selection technique is unnecessary in a large sample setting in which case a large number of mixture components is suggested.  
A set of numerical examples demonstrates that the proposed algorithm outperforms the standard iCE and the BiCE with the independent categorical distribution. 
Note that there is no guarantee that the BiCE-CM can find all major failure modes. 
The accuracy and efficiency of the BiCE-CM depend highly on the choice of the prior distribution. 
In this paper, we suggest a balanced prior that works well in all our numerical examples. 
A detailed investigation of alternative choices of the prior should be carried out. 
In addition, the BiCE-CM method does not directly apply to high dimensional problems due to the degeneration of the IS weights, and hence, dimensionality reduction techniques should be employed in such cases. 
These two aspects will be addressed in future work.

\section{Acknowledgment}
The first author gratefully acknowledges the financial support of the China Scholarship Council.

\appendix

\bibliographystyle{IeeeTran}
\bibliography{mybib}






\end{document}